\documentclass[prx,twocolumn,longbibliography,superscriptaddress,preprintnumbers]{revtex4-2}
\usepackage{comment}
\usepackage{bm}
\usepackage{dsfont}
\usepackage{graphicx}
\usepackage{physics}
\usepackage{epsfig}
\usepackage{amsfonts}
\usepackage{epstopdf}
\usepackage{balance}
\usepackage[dvipsnames]{xcolor}
\usepackage{calc}
\usepackage{natbib}
\usepackage[colorlinks,
            linkcolor=blue,
            anchorcolor=blue,
            citecolor=blue,
            urlcolor=blue]{hyperref}
\usepackage{lipsum}
\usepackage[version=3]{mhchem} 
\usepackage{color}
\usepackage{soul}
\usepackage[etex=true,export]{adjustbox}
\usepackage{makecell}
\usepackage{multirow}
\usepackage{tabularx,multirow,array,diagbox}
\usepackage{adjustbox}
\usepackage{booktabs}
\makeatletter
\renewcommand{\maketag@@@}[1]
{\hbox{\m@th\normalsize\normalfont#1}}%

\newcommand\bl[1]{\boldsymbol{#1}}
\newcommand\wt[1]{\widetilde{#1}}
\newcommand\ov[1]{\overline{#1}}

\newcommand\LLsr[2]{\Psi^{(#1)}_{#2,\boldsymbol{k}}(\boldsymbol{r})}
\newcommand\gLLr[1]{\Theta^{(s)}_{#1,\boldsymbol{k}}(\boldsymbol{r})}
\newcommand\gLLsr[2]{\Theta^{(#1)}_{#2,\boldsymbol{k}}(\boldsymbol{r})}
\newcommand\gLLu[1]{\widetilde{\Theta}^{(-)}_{#1,\boldsymbol{k}}}
\newcommand\gLLur[1]{\widetilde{\Theta}^{(-)}_{#1,\boldsymbol{k}}(\boldsymbol{r})}

\newcommand\e[1]{e^{(s)}_{#1,\boldsymbol{k}}}

\newcommand\er[1]{e^{(s)}_{#1,\boldsymbol{k}}(\boldsymbol{r})}

\makeatother
  
\begin{document}

\title{Abelian and Non-Abelian Fractionalized States in Twisted MoTe$_2$: A Generalized Landau-Level Theory}

\author{Bohao Li}
\affiliation{School of Physics and Technology, Wuhan University, Wuhan 430072, China}
\author{Yunze Ouyang}
\affiliation{School of Physics and Technology, Wuhan University, Wuhan 430072, China}
\author{Fengcheng Wu}
\email{wufcheng@whu.edu.cn}
\affiliation{School of Physics and Technology, Wuhan University, Wuhan 430072, China}
\affiliation{Wuhan Institute of Quantum Technology, Wuhan 430206, China}

\begin{abstract}
Fractional Chern insulators are lattice analogs of fractional quantum Hall states that realize fractionalized quasiparticles without an external magnetic field. A key strategy to understand and design these phases is to map Chern bands onto Landau levels (LLs). Here we introduce a universal framework that variationally decomposes Bloch bands into generalized LLs, providing a controlled and quantitative characterization of their effective LL nature. Applying this approach to twisted bilayer MoTe$_2$ modeled by first-principles-derived moir\'e Hamiltonians, we find that the first moir\'e valence band is dominated by the generalized zeroth LL across a broad range of twist angles. Exact diagonalization further supports the formation of Abelian fractional Chern insulators in the Jain sequences.  The second moir\'e band, renormalized via Hartree-Fock calculations at hole filling $\nu_h = 2$, is dominated by the generalized first LL at twist angles $\theta = 2.45^\circ$ and $2.13^\circ$. At $\theta = 2.45^\circ$, we find numerical evidence for a non-Abelian Moore--Read (MR) state at $\nu_h = 5/2$, with consistent signatures in both the energy spectrum and the particle entanglement spectrum. Interpolation studies further demonstrate an adiabatic connection between this state and the MR state in the conventional first LL. In contrast, at $\theta = 2.13^\circ$, a charge-density-wave state prevails in the competition with the MR state due to the larger bandwidth.  The approach of decomposing Bloch bands into generalized LLs offers a theoretical framework for investigating exotic fractionalized phases, including non-Abelian states, in realistic systems.
\end{abstract}
\maketitle

\section{introduction}
Fractionalized states are exotic quantum phases in which the elementary excitations are quasiparticles that carry fractional charge and obey anyonic statistics, which can be either Abelian or non-Abelian. A key example is the fractional quantum Hall insulators (FQHIs), which arise in Landau levels (LLs) of a two-dimensional electron system subjected to low temperatures and high perpendicular magnetic fields \cite{Tsui1982Two}. The theoretical foundation of these states was first established by Laughlin's trial many-body wavefunction \cite{Laughlin1983}, which describes fractionalized states at simple odd-denominator filling factors, such as the $\nu=1/3$ state in the zeroth LL (0LL), and features quasiparticle excitations that obey Abelian statistics. This framework was later extended to a broader range of odd-denominator filling factors using hierarchy schemes~\cite{Haldane1983Fractional, Halperin1984Statistics} and composite fermion theory \cite{ Jain1989Composite, Jain1990Theory}. In contrast, even-denominator FQHIs are less common but have been observed at filling factor $\nu=5/2$ in the first LL (1LL) \cite{Willett1987Observation}, where the Moore–Read (MR) state is the leading theoretical candidate \cite{Moore1991Nonabelions,Read1992Fractional,Read2000Paired}. The MR state is distinguished by non-Abelian quasiparticle excitations, which have attracted great interest due to their potential for topological quantum computation \cite{DasSarma2005Topologically}.

The concept of FQHIs has been extended to fractional Chern insulators (FCIs) in lattice model systems, where Chern bands replace the role of LLs \cite{Tang2011,Sun2011,Neupert2011,Regnault2011Fractional,Sheng2011Fractional}. 
Like their continuum counterparts, FCIs exhibit topological ground-state degeneracy on a torus and support fractionalized quasiparticle excitations. Crucially, FCIs can provide a new paradigm for realizing fractionalized phases without the need for an external magnetic field, emerging instead from spontaneous time-reversal symmetry breaking. The experimental breakthrough in realizing FCIs at zero external magnetic field was marked by the observation of the fractional quantum anomalous Hall effect in twisted bilayer MoTe$_2$ (tMoTe$_2$) \cite{Cai2023,Zeng2023,Park2023,Xu2023Observation} and subsequently in moir\'e rhombohedral multilayer graphene systems
\cite{lu2024fractional,Xie2025Tunable}. 
These experimental observations have not only validated the FCI concept but also spurred a surge of theoretical investigations.
These efforts encompass FCIs in the Jain sequence \cite{Reddy2023Toward,Reddy2023Fractional,Wang2024Fractional,Xu2024Maximally,mao2024lattice,yu2024fractional,Abouelkomsan2024,Song2024,Nicolas2024Magic,Lu2024FractionalChern}, fractionalized quasiparticle excitations \cite{Liu2025Characterization,Miguel2025Spinless}, anomalous composite Fermi liquids \cite{Goldman2023Zero,Dong2023,Reddy2023Toward}, as well as proposals for fractional topological insulators \cite{jian2024minimal} and non-Abelian phases \cite{zhang2024nonabelian,Ahn2024NonAbelian,maymann2025theory,Xu2025Multiple,Chen2025Robust,Wang2025Higher,reddy2024nonabelian}, which elucidates the microscopic mechanisms and emergent properties of these fractionalized phases.


A key theoretical strategy for understanding and designing  FCIs is to establish their connections with FQHIs through correspondence between Chern bands and LLs \cite{Qi2011Generic,Wu2012Gauge,Siddharth2013,Jackson2015,Claassen2015,Tarnopolsky2019,Ledwith2020,Wang2021Chiral,Wang2021Exact,Ozawa2021Relations,Mera2021Kahler,Wang2022Hierarchy,Ledwith2022IdealChern,Ledwith2023Vortexability,Wang2023Origin,Dong2023Manybody,Morales2024Magic,Shi2024Adiabatic,li2025Variational,Fujimoto2025Higher}. An important result is the identification of Chern bands with ideal quantum geometry that saturates the trace inequality, for which an exact correspondence between the Bloch wave function and the generalized 0LL has been established \cite{Wang2021Exact}. In this context, the generalized 0LL wave function is the conventional 0LL modulated by a position-dependent, momentum-independent function $\mathcal{B}(\bl{r})$.
Such ideal Chern bands can be theoretically realized in twisted bilayer (or multilayer) graphene in the chiral limit \cite{Tarnopolsky2019,Wang2021Chiral,Wang2022Hierarchy,Ledwith2022IdealChern}, as well as in the Aharonov-Casher model under imhomogenous magnetic field \cite{Aharonov1979Ground,Shi2024Adiabatic}.

In realistic materials, Chern bands inevitably deviate from the ideal limit, raising a fundamental question: to what extent, and in what sense, can lattice Chern bands be mapped onto LLs? Addressing this question is central to understanding why certain Chern bands support fractionalized topological phases analogous to those in the fractional quantum Hall effect, while others do not. Moir\'e materials provide a unique platform for tackling this problem, as their band structures are highly tunable and can approach the ideal Chern-band limit in certain cases. In tMoTe$_2$, the first moir\'e valence band closely approximates an ideal Chern band, a property attributed to the layer-pseudospin skyrmion texture \cite{Wu2019Topological} that generates an emergent magnetic field within the adiabatic approximation \cite{Morales2024Magic,Shi2024Adiabatic}. This band has been mapped to a generalized 0LL using a variational approach \cite{li2025Variational}. Beyond the lowest band, first-principles calculations indicate that the second moir\'e band at twist angles around $\theta \approx 2^{\circ}$ exhibits quantum geometric properties resembling those of the 1LL \cite{Wang2025Higher,Xu2025Multiple}. Consistently, exact diagonalization (ED) calculations reveal signatures of non-Abelian fractionalized states in this band, analogous to those found in the 1LL \cite{Ahn2024NonAbelian,Wang2025Higher,Xu2025Multiple,Chen2025Robust}. These results underscore the importance of establishing a quantitative mapping between Chern bands and LLs. Developing a general and systematic framework for such a correspondence is essential for elucidating the conditions under which Chern bands host Abelian and non-Abelian fractionalized phases, and remains a key open problem with implications for moiré systems and topological materials more broadly.

In this work, we develop a variational framework that decomposes realistic Bloch bands into generalized LLs. The generalized LLs with indices greater than zero are extended from the generalized 0LL, possessing a quantum geometry that is no longer ideal, while preserving a quantized integrated trace of the quantum metric \cite{liu2025theoryofgLL}. Crucially, these generalized LLs form a complete and orthonormal basis, enabling a controlled decomposition of realistic Bloch bands. By introducing a variational mapping approach, we establish a generalized LL representation of the Bloch bands, where the decomposition can be dominated by a specific generalized LL. This dominance is achieved by variationally optimizing the spatially varying function $\mathcal{B}(\mathbf{r})$, thereby maximizing the weight of the targeted generalized LL. This approach provides a concrete route to identifying the effective LL character of moir\'e Chern bands and to connecting their quantum geometry to the nature of the fractionalized phases they host.

We apply this variational mapping to  tMoTe$_2$ to investigate Abelian and non-Abelian fractionalized states. Our study uses a continuum model for tMoTe$_2$ in its most generic form, incorporating all symmetry-allowed terms and first-principles derived parameters \cite{zhang2024universal}. In this continuum model, the first moir\'e band in $+K$ valley have the Chern number $\mathcal C=1$ within the twist angle range of $2.13^\circ\le\theta\le 3.89^\circ$ under study. It nearly, though not exactly, saturates the trace inequality. The second moir\'e band  has the same Chern number as that of the first band within a smaller twist angle range \(2.13^\circ \leq \theta \leq 2.88^\circ\), and approaches the integrated trace condition of the generalized 1LL. We variationally decompose the first (second) moir\'e band into generalized LLs, which reveals that the generalized zeroth (first) LL is the dominant component, establishing a quantitative correspondence between moir\'e Chern bands and their effective LL counterparts.  We emphasize that even if a band’s quantum geometry satisfies the integrated trace condition of the generalized 1LL, this does not guarantee a corresponding resemblance in their wave functions. Our explicit decomposition is, therefore, crucial to reveal the internal generalized-LL structure of the Bloch bands.

We perform ED studies with Coulomb interactions projected onto both the original wave function and the variationally obtained generalized 0LL wave function for the first band. In both cases, the evidence of FCIs is found at hole filling factors $\nu_h=1/3,2/5,3/5$, and $2/3$ in the Jain sequences. At $\nu_h=2/3$, FCI remains stable within $2.13^\circ\le\theta\le 3.89^\circ$. At $\nu_h=1/3$, FCI remains stable in a narrower region $2.45^\circ\le\theta\le 3.48^\circ$, where the system becomes gapless at $\theta=2.13^\circ$ and transitions into a charge density wave (CDW) state at $\theta=3.89^\circ$. At $\nu_h=2/5$ and $3/5$, FCI remains stable within $2.45^\circ\le\theta\le 3.89^\circ$ and becomes gapless at $\theta=2.13^\circ$. The ED spectrum in both models has a good correspondence at the four filling factors, showing that these Abelian fractionalized states in tMoTe$_2$ are adiabatically connected to their corresponding FQHIs. Furthermore, FCI states are also identified at $\nu_h=4/7$ within $2.13^\circ\le\theta\le 3.89^\circ$, and $\nu_h=3/7, 5/9$, and $4/9$ within $2.45^\circ\le\theta\le 3.89^\circ$, which also belong to the Jain sequences. Numerical evidence for an anomalous composite Fermi liquid is observed at $\nu_h = 1/2$ across the studied range of twist angles.

We investigate the possible non-Abelian state in the second band of tMoTe$_2$, focusing on the filling factor $\nu_h = 5/2$ and twist angle $\theta=2.45^\circ$. We first perform a self-consistent Hartree-Fock (HF) calculation at the integer filling $\nu_h = 2$  to construct interaction-renormalized Bloch bands. In this refined basis, we also perform mapping between the second band and the generalized LLs, where the weight of the generalized 1LL is further enhanced compared to the noninteracting case. Experimental results show a ferromagnetic state near $\nu_h=5/2$ and $\nu_h=3$ at various twist angles \cite{Xu2025Interplay,Park2025Ferromagnetism,An2025Observation}, indicating a trend of spontaneous valley polarization in the second moir\'e band. Therefore, we use the renormalized Bloch states to build a projected many-body Hamiltonian at $\nu_h = 5/2$ within the second band and $+K$ valley. We then perform ED calculation, where both the ED spectrum and the particle entanglement spectrum (PES) show consistent evidence of the non-Abelian MR state. Additionally, we perform ED studies using a series of Hamiltonians interpolating between the second band of tMoTe$_2$ and the 1LL. We find a finite gap throughout, indicating that the MR state is adiabatically connected to the corresponding state in the 1LL. At a smaller twist angle, $\theta = 2.13^\circ$, although the second band also exhibits a high generalized 1LL weight, ED and PES indicate a charge-density-wave (CDW) state, which is favored over the MR state due to the larger bandwidth.

The rest of this paper is organized as follows. In Section \ref{sec:review_gLL}, we provide a review of generalized LLs and present their quantum geometric properties. Section \ref{sec:Hamiltonian} is devoted to the continuum model for tMoTe$_2$, along with a detailed analysis of the first two moir\'e valence bands across different twist angles. In Section \ref{sec:mapping}, we introduce a variational mapping method to construct a generalized LL representation of Bloch bands, which is then applied to the first two bands of tMoTe$_2$. Section \ref{sec:Abelian_FCI} investigates Abelian fractional Chern insulators in the first moir\'e band. Section \ref{sec:nonAbelian_FCI} presents the numerical evidence of the MR state in the second moir\'e band at half filling. Finally, in Section \ref{sec:Discussion} we conclude with a summary and a discussion. Additional technical details and numerical results are provided in appendices.

\section{generalized LLs}
\label{sec:review_gLL}
We begin by reviewing the concept of quantum geometry and introducing the generalized LL wave functions.  These generalized LLs extend the standard LL wave function by incorporating spatial modulations and fluctuations in quantum geometry \cite{liu2025theoryofgLL}.  As a starting point, we present the definition of the quantum geometric tensor (QGT),
\begin{equation}
\label{QGT_definition}
\begin{aligned}
(\chi_{\bl k})_{ab}=\bra{\partial_{k_a}u_{\bl k}}(\mathds{1}-P)\ket{\partial_{k_b}u_{\bl k}}.
\end{aligned}     
\end{equation}
Here we focus on the Abelian QGT for a single band of Bloch wave function $\psi_{\bl k}(\bl r)$ with $\bl k$  the wave vector. In Eq.~\eqref{QGT_definition}, $u_{\bl k}(\bl r)=e^{-i\bl k\cdot\bl r}\psi_{\bl k}(\bl r)$ is the periodic part of $\psi_{\bl k}(\bl r)$
, $P=\ket{u_{\bl k}}\!\bra{u_{\bl k}}$ is the projection operator, and $a ,b $ denote spatial indices. We focus on two dimensions here.  The Berry curvature $\Omega_{\bl k}$ and quantum metric $g_{\bl k}$ are derived from Eq.~\eqref{QGT_definition} through the relation $(\chi_{\bl k})_{ab}=(g_{\bl k})_{ab}-\frac{i}{2}\epsilon_{ab}\Omega_{\bl k}$, where $\epsilon_{ab}$ is the Levi-Civita symbol. Here $\chi_{\bl 
k}$ is a Hermitian matrix, $g_{\bl k}$ is a real symmetric matrix, and $\Omega_{\bl k}$ is a real function. Because $\mathds{1}-P$ is a projector, the QGT is semipositive definite, which implies that $\det [g_{\bl k}] \geq \Omega_{\bl k}^2/4$. Since $g_{\bl k}$ is a $2\times 2$ real symmetric matrix, $(\text{Tr} [g_{\bl k}])^2 \geq 4 \det [g_{\bl k}]$. Therefore, there is a trace inequality $\mathrm{Tr}[g_{\bl k}] \geq \abs{\Omega_{\bl k}}$.

The integrals of the Berry curvature and the trace of quantum metric over the Brillouin zone give rise to, respectively, the Chern number $\mathcal{C}$ and the quantum weight $\mathcal{K}$ \cite{Onishi2025Quantum},
\begin{equation}
\begin{aligned}
    \mathcal{C}=\frac{1}{2\pi} \int d {\bm k} \,\Omega_{\bm k}, \\
    \mathcal{K}=\frac{1}{2\pi} \int d {\bm k} \,\text{Tr} [g_{\bm k}]. \\
\end{aligned}    
\end{equation}
Here, $\mathcal{C}$ is a topological invariant that is quantized to an integer value, whereas $\mathcal{K}$ is not. 
Because of the inequality $\mathrm{Tr}[g_{\bl k}] \geq \abs{\Omega_{\bl k}}$, we have $\mathcal{K} \geq |\mathcal{C}|$, which represents an alternative form of the trace inequality.
As discussed below, both $\mathcal{C}$ and $\mathcal{K}$ provide important insights into the quantum geometry, complementing the information captured by the QGT.  

A prototypical example of topological bands with finite Chern numbers is the LLs formed in two-dimensional electron gas  subjected to a perpendicular magnetic field. The wave function of LLs satisfying magnetic translational symmetry is the magnetic Bloch wave function given by $\Psi_{n,\bl k}^{(s)}(\bl r)$ \cite{Haldane2018modular}, where $n \geq 0$ is the LL index, $\bl k$ is the magnetic translation quantum number and $s=\pm$ denote the orientation of magnetic field. 
We present the form of $\Psi_{n,\bl k}^{(s)}(\bl r)$ in Appendix~\ref{appendix:A}.
The QGT for the $n$th LL ($n$LL), with wave function $\Psi_{n,\bl k}^{(s)}(\bl r)$, is
\begin{equation}
\label{QGT-nLL}
\begin{aligned}
\chi_{\bl k, n, s}^{(\mathrm{LL})}=\begin{pmatrix}
n+\frac{1}{2} & -\frac{is}{2} \\
\frac{is}{2} & n+\frac{1}{2}
\end{pmatrix}\ell^2,
\end{aligned}     
\end{equation}
where $\ell$ is the magnetic length. The magnetic unit cell, which contains one flux quantum, has an area of $2\pi \ell^2$.
The QGT for the $n$LL is momentum-independent, characterized by $\mathrm{Tr}[g_{\bl k}] = (2n + 1)\ell^2$ and $\abs{\Omega_{\bl k}} = \ell^2$. These satisfy the relation $\mathrm{Tr}[g_{\bl k}] = (2n + 1)\abs{\Omega_{\bl k}}$ at each momentum $\bl k$. The $0$LL is special in that it saturates the trace inequality $\mathrm{Tr}[g_{\bl k}] \geq \abs{\Omega_{\bl k}}$. Because $\abs{\Omega_{\bl k}} = \ell^2$ is independent of the LL index $n$, each LL has a Chern number with a magnitude $|\mathcal{C}|=1$. However, the quantum weight $\mathcal{K}$ depends on the LL index $n$, given by $\mathcal{K}=2n+1$.

The property of the $0$LL motivates the construction of more general Bloch states that also saturate the trace inequality, which are generalized $0$LL with wave function given by
\begin{equation}
\label{g0LL}
\begin{aligned}
\Theta_{0,\bl k}^{(s)}(\bl r)=\mathcal{N}_{0, \bl k}\mathcal B^{(s)}(\bl r)\LLsr{s}{0}.
\end{aligned}     
\end{equation}
Here $\mathcal{B}^{(s)}(\bl r)$ is a position $\bl r$-dependent but momentum $\bl k$-independent function, and $\mathcal{N}_{0,\bl k}$ is a normalization factor. Due to the spatial modulation introduced by $\mathcal{B}^{(s)}(\bl r)$, the QGT of $\Theta_{0,\bl k}^{(s)}(\bl r)$ acquires momentum dependence. Nevertheless, it still satisfies the equality $\mathrm{Tr}[g_{\bl k}] = \abs{\Omega_{\bl k}}$ at each momentum $\bl k$, a condition known as ideal quantum geometry \cite{Wang2021Exact}. The Chern number and quantum weight associated with $\Theta_{0,\bl k}^{(s)}(\bl r)$ remain the same as those of the $0$LL, given by $\mathcal{K}=|\mathcal{C}|=1$. An important feature of the generalized $0$LL described by $\Theta_{0,\bl k}^{(s)}(\bl r)$ is that it enables construction of trial wave functions for fractional Chern insulators (FCIs) \cite{Ledwith2020,Wang2021Exact}. The trial state takes the following form 
\begin{equation}
    \Phi_F = \Psi_F \prod_i \mathcal{B}^{(s)}(\bl r_i), 
\label{PhiF}
\end{equation}
where $\bl r_i$ denotes the position of the $i$th electron, and $\Psi_F$ is a fractional quantum Hall state in the 0LL. In certain cases, $\Phi_F$ can serve as the exact ground state of short-range repulsive interactions at special filling factors \cite{Wang2021Exact}.

Recent theoretical developments show that the generalized $n$LL wave function $\Theta_{n,\bl k}^{(s)}(\bl r)$ for $n \geq 1$ can be systematically constructed by applying Gram–Schmidt orthogonalization to a set of density-modulated basis functions $e_{n,\bl k}^{(s)}(\bl r)$ \cite{liu2025theoryofgLL}, defined as,
\begin{equation}
\label{enLL}
\begin{aligned}
e_{n,\bl k}^{(s)}(\bl r)=\mathcal B^{(s)}(\bl r)\Psi_{n,\bl k}^{(s)}(\bl r).
\end{aligned}     
\end{equation}
Here $\mathcal{B}^{(s)}(\bl r)$ introduces spatial modulation and $\Psi_{n,\bl k}^{(s)}(\bl r)$ is the magnetic Bloch wave function of the conventional $n$LL. The generalized LL wave function $\Theta_{n,\bl k}^{(s)}(\bl r)$ is then given by,
\begin{equation}
\begin{small}
\label{gnLL}
\begin{aligned}
\gLLr{n} = 
\begin{cases}
\mathcal N_{0,\bl k} \er{0} &\!\! n = 0\\
\mathcal N_{n,\bl k}[ \er{n} - \sum\limits_{m=0}^{n-1} \langle \Theta_{m,\bl k}^{(s)} | \e{n} \rangle \Theta_{m,\bl k}^{(s)}(\bl r)]& \!\!n \geq 1,
\end{cases}
\end{aligned}
\end{small}
\end{equation}
where $\mathcal N_{n,\bl k}$ is the normalization factor. In the case of $\mathcal B^{(s)}(\bl r)=1$, $\Theta_{n,\bl k}^{(s)}(\bl r)$ reduces to $\Psi_{n,\bl k}^{(s)}(\bl r)$. For $n \geq 1$, the QGT of $\gLLr{n}$ is generally momentum dependent and does $\textit{not}$ obey the local trace condition of $\mathrm{Tr}[g_{\bl k}] = (2n+1) \abs{\Omega_{\bl k}}$, but follows the integrated form of the trace condition (see Ref.~\cite{liu2025theoryofgLL} and also Appendix~\ref{appendix:B} for proof),
\begin{equation}
\begin{aligned}
\label{trace-condition2}
\mathcal{K}=(2n+1)\abs{\mathcal C}=2n+1.
\end{aligned}     
\end{equation} 
We emphasize that while a band with $\mathcal{K}=|\mathcal{C}|=1$ admits a universal wave function—the generalized $0$LL given in Eq.~\eqref{g0LL}—this universality breaks down for higher indices. Specifically, for $n>0$, even if a band satisfies Eq.~\eqref{trace-condition2}, it does not necessarily realize the generalized $n$LL as its wave function.
 
The generalized $n$LL wave functions $\Theta_{n,\bl k}^{(s)}(\bl r)$ can form a complete and orthonormal basis for decomposing Bloch bands. In this work, we focus on Chern bands in tMoTe$_2$, where the wave functions are expressed in the layer pseudospin$-1/2$ space. Therefore, we consider a two-component $\mathcal{B}^{(s)}(\bl r)$ to account for the layer degree of freedom. Moreover, we require that the two-component spinors $\mathcal{B}^{(+)}(\bl r)$ and $\mathcal{B}^{(-)}(\bl r)$ are orthogonal at each position $\bl r$, which ensures the following orthonormal condition,
\begin{equation}
    \label{ortho_cond}
    \langle \Theta_{n,\bl k}^{(s)} | \Theta_{n',\bl k}^{(s')} \rangle = \delta_{n,n'} \delta_{s,s'}.
\end{equation}
In our work, we further require that $\Theta_{n,\bl k}^{(s)}(\bl r)$ obeys Bloch theorem instead of magnetic Bloch theorem, which puts constrains on the translational symmetry of $\mathcal{B}^{(s)}(\bl r)$ (see Appendix~\ref{appendix:A} for discussion).

\begin{figure}[t]
    \includegraphics[width=1.\columnwidth]{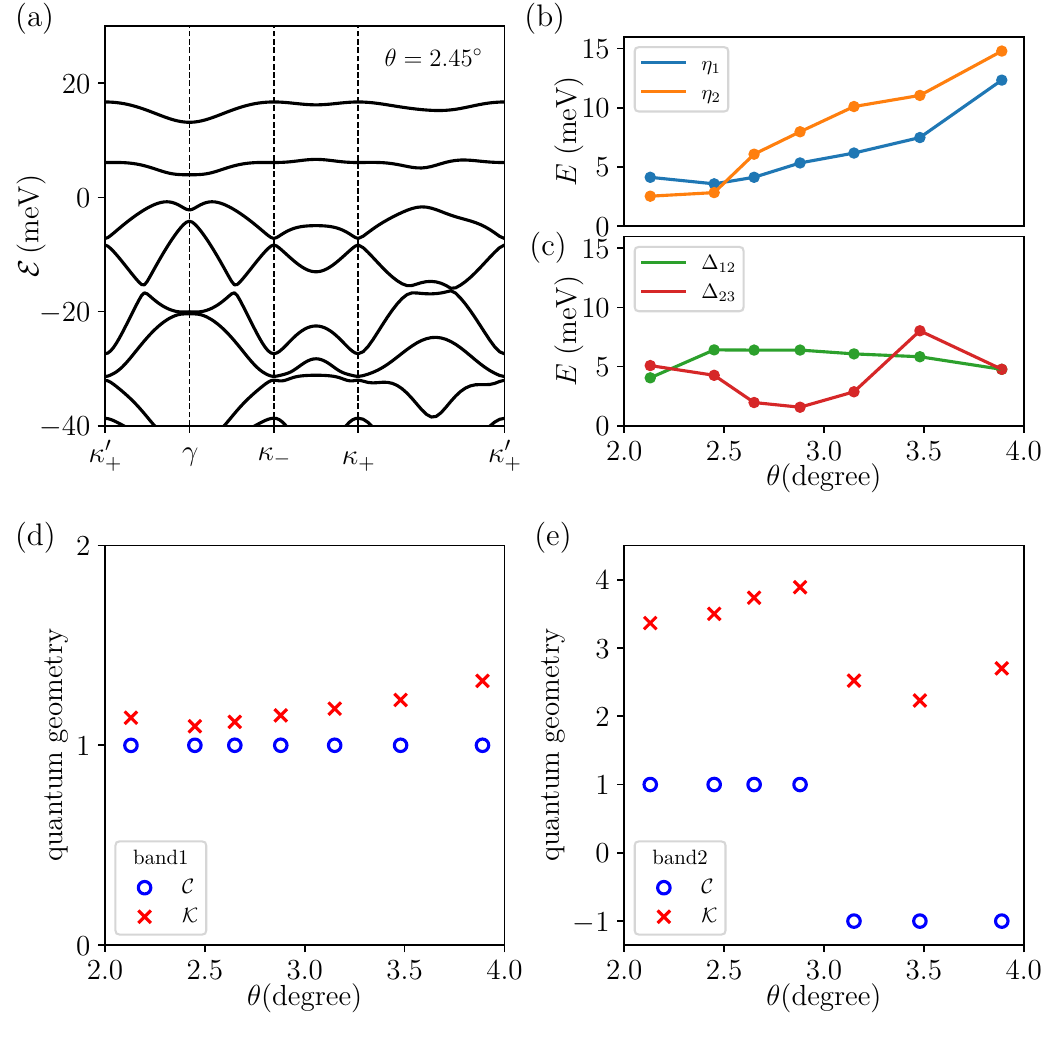}
    \caption{(a) Moir\'e band structure of tMoTe$_2$ in $+K$ valley at $\theta=2.45^\circ$ along high-symmetry path in the mBZ. (b) Bandwidths of the first moir\'e band $\eta_1$ and the second moir\'e band $\eta_2$ as functions of $\theta$. (c) Energy gaps $\Delta_{12}$ between the first and second bands, and $\Delta_{23}$ between the second and third bands,  plotted as a function of $\theta$. (d) Chern number $\mathcal{C}$ and quantum weight $\mathcal{K}$ for the first moir\'e valence band in $+K$ valley. (e) Same as (d), but for the second moir\'e band. Results are shown for commensurate angles $\theta$ at $2.13^\circ, 2.45^\circ, 2.65^\circ, 2.88^\circ, 3.15^\circ, 3.48^\circ$ and $3.89^\circ$.} 
    \label{fig:band1}
\end{figure}

\section{moir\'e bands}
\label{sec:Hamiltonian}
In $t$MoTe$_2$, the single-particle moir\'e Hamiltonian for valence-band states can be constructed independently for the $+K$ and $-K$ valleys. We focus on the $+K$ valley, as the $-K$-valley Hamiltonian follows from it via time-reversal symmetry $\mathcal{T}$. The continuum moir\'e Hamiltonian is typically represented as a $2 \times 2$ matrix in the layer-pseudospin basis, incorporating quadratic kinetic terms as well as the lowest-harmonic intralayer potentials and interlayer tunneling terms \cite{Wu2019Topological}. In this work, we adopt the moir\'e Hamiltonian in the most generic form following Ref.~\cite{zhang2024universal}, which includes (i) generalized kinetic-energy terms beyond the quadratic approximation and (ii) a momentum-dependent potential terms beyond the lowest-harmonic approximation. The $+K$-valley Hamiltonian is then given by
\begin{equation}
 \label{H0}
 \begin{small}
 \begin{aligned}
 H_+=U_0 (\bl r)\begin{pmatrix}
H^0_b(\hat{\bl{k}})+\Delta_{b}(\hat{\bl{k}},\bl r) & \Delta_T(\hat{\bl{k}},\bl r) \\
\Delta_T^\dagger(\hat{\bl{k}},\bl r) & H^0_t(\hat{\bl{k}})+\Delta_{t}(\hat{\bl{k}},\bl r)
\end{pmatrix}U_0^\dagger (\bl r),
 \end{aligned}     
 \end{small}
 \end{equation}
where $H_l^0(\hat{\bl k})$ denotes the layer-dependent kinetic term,  $\Delta_l(\hat{\bl k},\bl r)$ is the momentum-dependent intralayer moir\'e potential, $\Delta_T(\hat{\bl k},\bl r)$ the momentum-dependent interlayer tunneling term, and $l=b,t$ labels the layers. The unitary matrix $U_0 (\bl r) = \mathrm{diag}(e^{i\bl\kappa_+ \cdot\bl r}, e^{i\bl\kappa_- \cdot\bl r} )$ is employed to render the Hamiltonian in a more symmetric and compact form, where $\boldsymbol{\kappa}_\pm=\frac{4\pi}{3a_M}(-\frac{\sqrt{3}}{2},\mp\frac{1}{2})$ are located at corners of moir\'e Brillouin zone (mBZ), $a_M \approx a_0/\theta$ is the moir\'e period, $\theta$ is the twist angle, and $a_0=3.52 ~\text{\AA}$ is the monolayer lattice constant. $H_l^0(\hat{\bl k})$, $\Delta_l(\hat{\bl k},\bl r)$, and $\Delta_T(\hat{\bl k},\bl r)$ can be expressed as a series expansion in the momentum operator $\hat{\bl{k}}$ and a harmonic expansion in the position operator $\bm{r}$. 

The kinetic energy $H_l^0(\hat{\bl k})$ is parametrized as,
 \begin{equation}
 \label{T0}
 \begin{aligned}
H_l^0(\hat{\bl k})= & \sum_{M_1,M_2}\frac{\hbar^2}{2m^{(M_1,M_2)}_l}f^{(M_1,M_2)}_{\hat{\bl k}},
 \end{aligned}     
 \end{equation}
where $f_{\hat{\bl k}}^{(M_1,M_2)}=(\hat k_x+i\hat k_y)^{M_1}(\hat k_x-i\hat k_y)^{M_2}$  forms a polynomial basis in $\hat{\bl{k}}$, $M_{1,2}\in\mathbb N_0$ and $\hbar^2/(2m^{(M_1,M_2)}_l)$ is the corresponding expansion coefficient. Equation~\eqref{T0} reduces to the conventional parabolic kinetic term when $m^{(M_1,M_2)}_l$ is infinite for all $(M_1,M_2)\ne(1,1)$. 

The intralayer potential $\Delta_l(\hat{\bl k},\bl r)$ takes the form
 \begin{equation}
 \label{V-intra}
 \begin{small}
 \begin{aligned}
\Delta_l(\hat{\bl k},\bl r)\!=\!& \sum_{M_1,M_2,\bl g}V^{(M_1,M_2)}_{\bl g,l}f^{(M_1,M_2)}_{\hat{\bl k}}e^{i\bl g\cdot\bl r}+\mathrm{h.c.}\;,\\
 \end{aligned}     
 \end{small}
 \end{equation}
 where $V^{(M_1,M_2)}_{\bl g,l}$ is the expansion coefficient, $\bl g$ is the moir\'e reciprocal lattice vector defined as $\bl g=m\bl g_1+n\bl g_2$ with $m,n\in\mathbb Z$ and $\bl g_j= \frac{4\pi}{\sqrt{3}a_M}[\cos\frac{(j-1)\pi}{3},\sin\frac{(j-1)\pi}{3}]$.

The interlayer tunneling term $\Delta_T(\hat{\bl k},\bl r)$ is given by 
  \begin{equation}
 \label{V-inter}
 \begin{small}
 \begin{aligned}
 &\Delta_T(\hat{\bl k},\bl r) = \\
 &\sum_{M_1,M_2,\bl q}w_{\bl q}^{(M_1,M_2)}[f_{\hat {\bl k}}^{(M_1,M_2)}e^{-i\bl q\cdot\bl r}+e^{i(\hat R_{2x}\bl q)\cdot\bl r}f_{\hat {\bl k}}^{(M_2,M_1)}],
 \end{aligned}     
 \end{small}
 \end{equation}
where $w_{\bl q}^{(M_1,M_2)}$ is the expansion coefficient, $\hat R_{2x}$ is the two-fold rotational operator around $x$-axis, $\bl q=\bl g+\bl q_1$ and $\bl q_1=\bl\kappa_+-\bl \kappa_-$. $\Delta_l(\hat{\bl k},\bl r)$ and $\Delta_T(\hat{\bl k},\bl r)$ become momentum-independent when $V^{(M_1,M_2)}_{\bl g,l}=0$ and $w^{(M_1,M_2)}_{\bl q}=0$ for $(M_1,M_2)\ne(0,0)$.

The point group symmetry of tMoTe$_2$ is $D_3$, which includes the $C_{3z}$ and $C_{2y}$ symmetries. Here, $C_{nj}$ represents an $n$-fold rotation about the $j$-axis. The valley-projected Hamiltonian in Eq.~\eqref{H0} is required to be Hermitian and to remain invariant under the $C_{3z}$ and  $C_{2y}\mathcal T$ symmetries. As a consequence, the parameters in Eqs. \eqref{T0},\eqref{V-intra}, and \eqref{V-inter} are constrained by the following conditions
  \begin{equation}
 \label{paras}
 \begin{aligned}
 & m^{(M_1,M_2)}_l=[m^{(M_2,M_1)}_l]^*=[m^{(M_1,M_2)}_{-l}]^*,\\
 & V^{(M_1,M_2)}_{\bl g,l}=\omega^{M_1-M_2}V^{(M_1,M_2)}_{\hat R_3\bl g,l}, V^{(M_1,M_2)}_{\bl g,l}=[V^{(M_1,M_2)}_{\hat R_{2x}\bl g,-l}]^*, \\
 & w_{\bl q}^{(M_1,M_2)}= \omega^{M_1-M_2}w_{\hat R_3\bl q}^{(M_1,M_2)},
 \end{aligned}     
 \end{equation}
where $\omega=e^{i2\pi /3}$ and $m^{(M_1,M_2)}_l$ is finite only if $M_1-M_2\equiv 0\,(\text{mod}\;3)$.

A first-principles-based framework for constructing the general moir\'e Hamiltonian, free of any empirical fitting, was developed in Ref.~\cite{zhang2024universal}. Using this approach, the $\theta$-dependent model parameters for several commensurate twist angles within the range $\theta \in [2.13^\circ, 3.89^\circ]$ were obtained. In the present work, we adopt the parameters from the full continuum model constructed in Ref.~\cite{zhang2024universal} (the corresponding parameters are extracted from Ref.~\cite{zhang2024github}), which achieves energy deviations below 0.5 meV and wavefunction overlaps exceeding 97\% for the four highest-energy bands, compared to the first-principles results in the standard basis of pseudo-atomic orbital functions.

A representative moir\'e band structure in $+K$ valley at $\theta=2.45^\circ$ is shown in Fig.~\ref{fig:band1}(a). The top two moir\'e bands are energetically isolated and exhibit a narrow bandwidth. Figure~\ref{fig:band1}(b) shows the bandwidth $\eta_1$ and $\eta_2$ of the first (i.e., topmost) and second moir\'e bands, respectively. Both $\eta_{1}$ and $\eta_{2}$ are relatively small, below 5 meV for $\theta \leq 2.45^\circ$, and stay under 15 meV for $\theta$ up to $3.89^\circ$, indicating the bands retain a notably flat character over this range. We further plot the band gap $\Delta_{n(n+1)}$  between the $n$th and $(n+1)$th moir\'e valence bands in Fig.~\ref{fig:band1}(c). For the plotted range of $\theta$, $\Delta_{12}$ remains finite, on the order of 5 meV.  However, $\Delta_{23}$ has a minimum near $\theta=3^\circ$, implying a topological phase transition in the second moir\'e band as explained below.

To characterize the quantum geometry of the first two bands, we plot the Chern number $\mathcal C$ and quantum weight $\mathcal{K}$ as  functions of $\theta$ in Figs.~\ref{fig:band1}(d) and \ref{fig:band1}(e). For the first moir\'e valence band, $\mathcal C$ is quantized to $+1$ within $\theta\in[2.13^\circ,3.89^\circ]$, and $\mathcal{K}$ exceeds $|\mathcal C|$ as expected, with only a small difference between them. The $\mathcal{K}-|\mathcal{C}|$ value, which quantifies the deviation from ideal quantum geometry, remains below 0.5 across the entire $\theta$ range and reaches a minimum of 0.095 at $\theta=2.45^\circ$, suggesting a strong connection between the first moir\'e valence band and the generalized 0LL. 

In contrast, the second moir\'e valence band exhibits a Chern number  $\mathcal{C}=+1$ at small twist angles but undergoes a sign reversal to $-1$ above a critical angle $\theta_c$ between $2.88^\circ$ and $3.15^\circ$.  For $\theta < \theta_c$, $\mathcal{K}$ is close to $3|\mathcal C|$. Motivated by Eq.~\eqref{trace-condition2}, the quantity $\mathcal{K}-3|\mathcal{C}|$ serves as a measure of similarity between the second moir\'e band and the generalized 1LL. This difference decreases with decreasing $\theta$ below $\theta_c$ and reaches a minimum of 0.37 at $\theta = 2.13^\circ$, indicating a potential correspondence between the second moir\'e band and the generalized 1LL.

\begin{figure}[t]
    \includegraphics[width=1.\columnwidth]{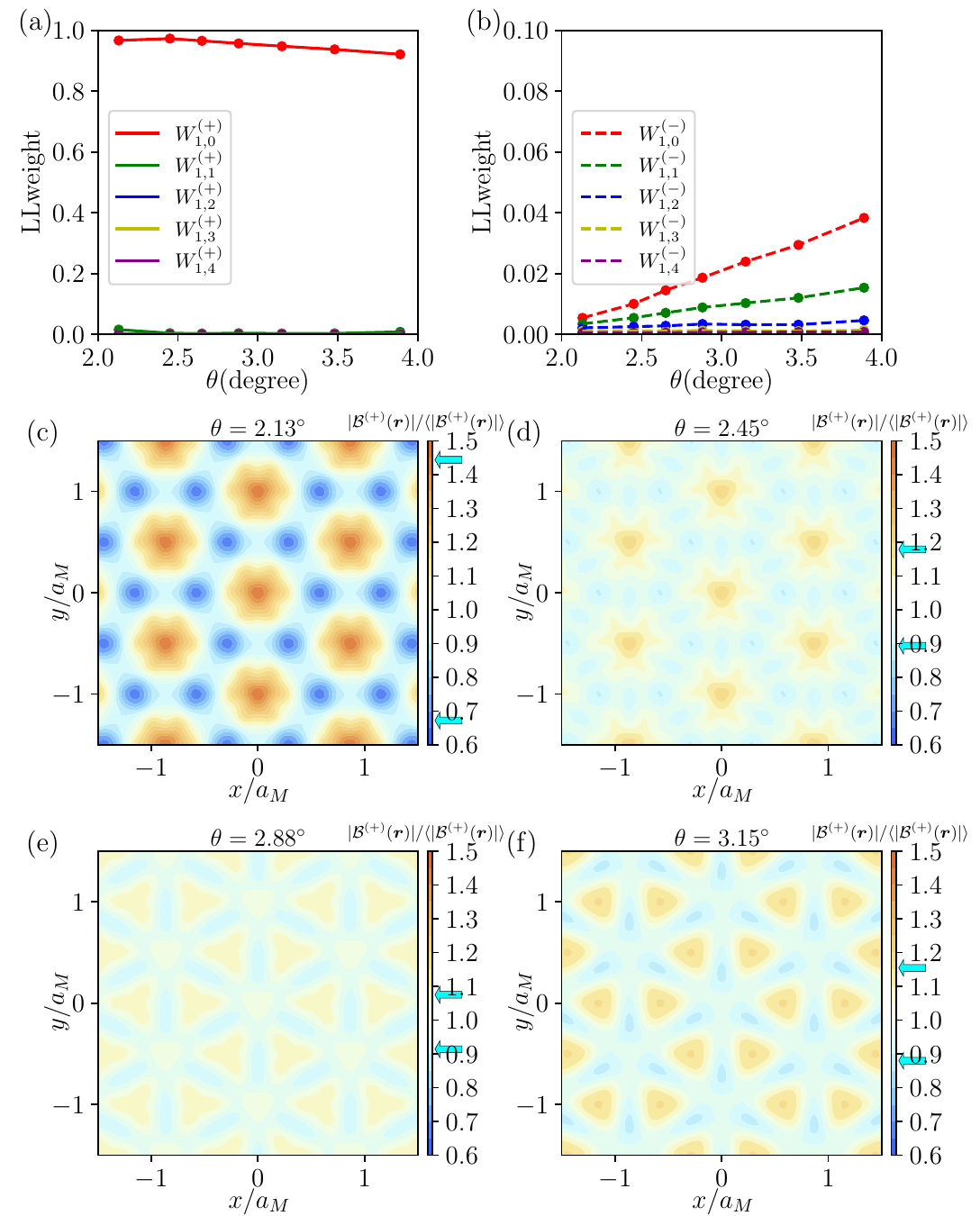}
    \caption{(a-b) LL weight $W_{0,n}^{(s)}$ for the first moir\'e band where $n$ and $s=\pm$ label, respectively, the index and chirality of the generalized LL. (c-f) Map of $\abs{\mathcal{B}^{(+)}(\bl r)}$ scaled by its spatial average at $\theta=2.13^\circ,2.45^\circ,2.88^\circ$, and $3.15^\circ$. The arrows indicate the data range in each plot.} 
    \label{fig:variation1}
\end{figure}

\begin{figure}[t]
    \includegraphics[width=1.\columnwidth]{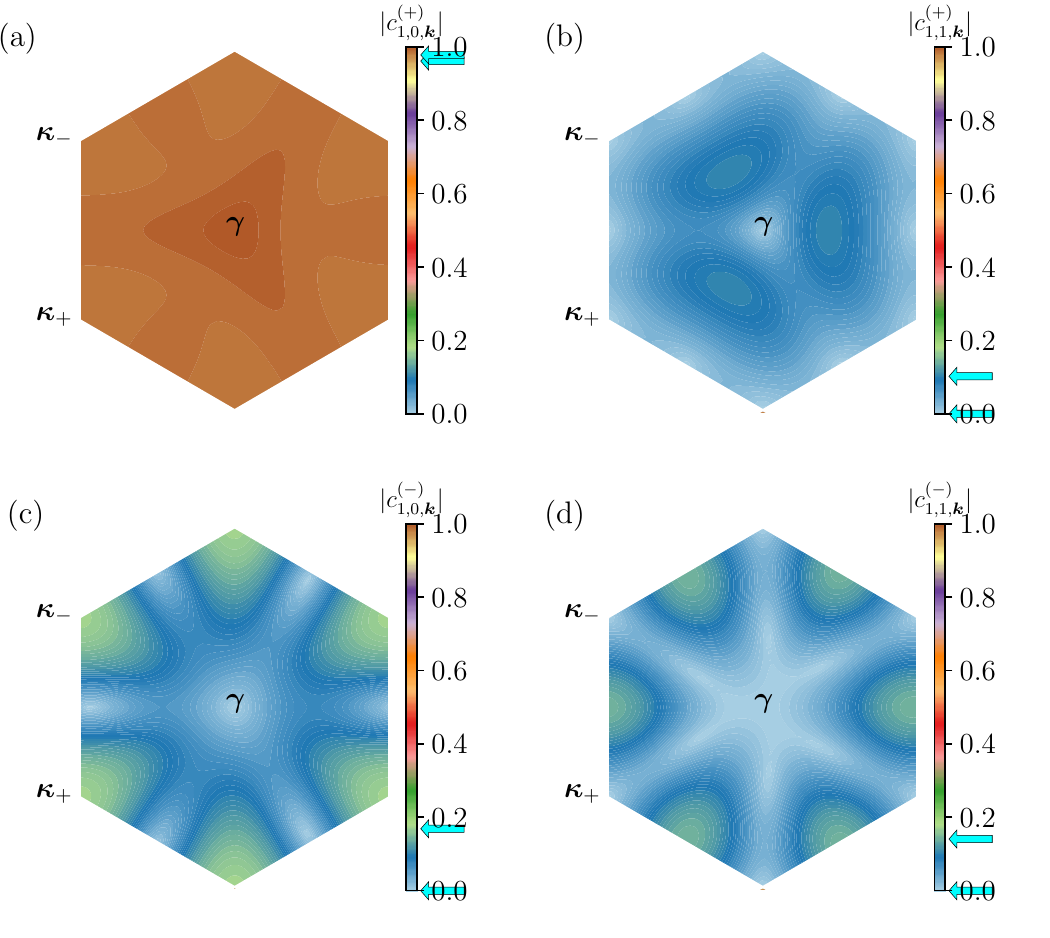}
    \caption{(a-d) Overlap $\lvert c_{1,n,\bl{k}}^{(s)}\rvert=\lvert\langle\Theta^{(s)}_{n,\bl k}\lvert\psi_{+,1,\bl k}\rangle\rvert$ at $\theta=2.45^\circ$ for $s=\pm$ and $n=0,1$ in the mBZ. The arrows indicate the data range in each plot.} 
    \label{fig:overlap_band1}
\end{figure}

\section{variational mapping} 
\label{sec:mapping}

We now introduce a variational mapping approach to establish a generalized LL representation of Bloch bands, and apply it to the first two moir\'e bands in the tMoTe$2$ model. The Bloch wave function $\psi_{+,m,\bl k}(\bl r)$ of the $m$th band at momentum $\bl k$ and $+K$ valley can be decomposed in terms of the generalized LL wave functions $\gLLr{n}$,
\begin{align}
\label{gLLdecomp}
\psi_{+,m,\bl  k}(\bl r)=  & \sum_{n}\sum_{s=\pm} c_{m,n,\bl k}^{(s)}\gLLr{n},
\end{align}
where $c_{m,n,\bl{k}}^{(s)}$ are the expansion coefficient. 
This decomposition is general as long as $\psi_{+,m,\bl k}(\bl r)$ and $\gLLr{n}$ share the same translational symmetry, in accordance with Bloch's theorem. To ensure this, we require that the magnetic Bloch wave function $\Psi_{n,\bl k}^{(s)}(\bl r)$, used in the definition of $\gLLr{n}$, is defined with respect to a magnetic unit cell that matches exactly with the moir\'e unit cell of tMoTe$_2$. This give the relation between the magnetic length $\ell$ and the moir\'e period $a_M$, $2\pi\ell^2=\sqrt{3}a_M^2/2$.

Using  the orthonormality condition in Eq.~\eqref{ortho_cond}, $c_{m,n,\bl{k}}^{(s)}$ can be expressed as the overlap between $\psi_{+,m,\bl k}(\bl r)$ and $\gLLr{n}$,  $c_{m,n,\bl{k}}^{(s)}=\langle\Theta^{(s)}_{n,\bl k}\lvert\psi_{+,m,\bl k}\rangle$. To measure the average overlap between $\psi_{+,m,\bl k}(\bl r)$ and $\gLLr{n}$, we define the generalized LL weight,
\begin{align}
W_{m,n}^{(s)}=\frac{1}{N}\sum_{\bl k}\lvert  c_{m,n,\bl{k}}^{(s)}\rvert^2,
\end{align}
where $N$ is the number of $\bl k$ points summed within the mBZ.

The spatially varying functions $\mathcal{B}^{(s)}(\bl r)$ are so far arbitrary, provided they respect the necessary translational symmetry to ensure that $\gLLr{n}$ is a Bloch wave function. Therefore, the decomposition in Eq.~\eqref{gLLdecomp} is not unique. For the decomposition to yield physical insight, we would require that a specific generalized LL, $\Theta_{n_0,\bl k}^{(s_0)}(\bl r)$, contributes a dominant weight $W_{m,n_0}^{(s_0)}$, compared to all other $W_{m,n}^{(s)}$. This can ensure that the dominant contribution to the wave function comes from a well-defined set of states, simplifying the physical interpretation and enabling insights from LL physics. The choice of $n_0$ and $s_0$ can be motivated by the quantum geometric properties of the targeted Bloch wave function $\psi_{+,m,\bl k}(\bl r)$ as explained below. We can then maximize $W_{m,n_0}^{(s_0)}$ variationally by adjusting $\mathcal{B}^{(s_0)}(\bm{r})$ via gradient ascent \cite{li2025Variational}, starting from an initial ansatz and iterating until convergence:
\begin{equation}
\label{iter}
\begin{aligned}
\mathcal{B}_i^{(s_0)}(\bl r)\rightarrow&
\mathcal{B}_i^{(s_0)}(\bl r)+\zeta \frac{\delta W_{m,n_0}^{(s_0)}}{\delta \ov{\mathcal{B}_i}^{(s_0)}(\bl r)},
\end{aligned}    
\end{equation}
where the subscript $i=1,2$ labels the $i$th component of the spinor $\mathcal{B}^{(s_0)}(\bl r)$, and $\zeta$ is a positive parameter. Here the functional derivative is defined as 
\begin{equation}
\begin{aligned}
\frac{\delta f}{\delta \ov g}=\frac{1}{2}[\frac{\delta f}{\delta \Re [g]}+i\frac{\delta f}{\delta \Im [g]}],
\end{aligned}
\end{equation} 
for a real function $f$.

\begin{figure}[t]
    \includegraphics[width=1.\columnwidth]{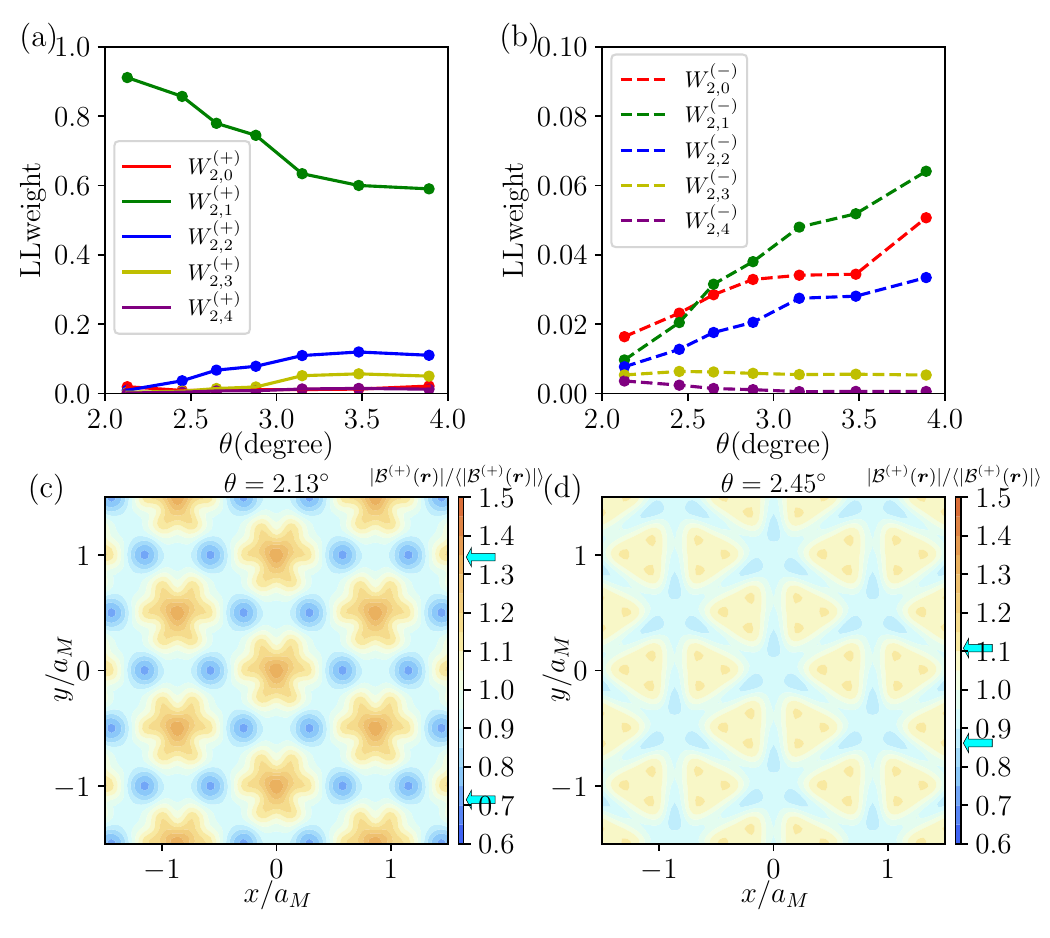}
    \caption{(a-b) LL weight $W_{2,n}^{(s)}$ for the second moir\'e band where $n$ and $s=\pm$ label, respectively, the index and chirality of the generalized LL. (c-d) Map of $\abs{\mathcal{B}(\bl r)}$ scaled by its spatial average at $\theta=2.13^\circ$ and $2.45^\circ$. The arrows indicate the data range in each plot.} 
    \label{fig:band2}
\end{figure}

We can further determine $\mathcal{B}^{(-s_0)}(\bl r)$ by noting it is a two-component spinor orthogonal to $\mathcal{B}^{(s_0)}(\bl r)$ at every $\bl r$. With this set of procedures, we can calculate all the coefficients $c_{m,n,\bl{k}}^{(s)}$ and the weights $W_{m,n}^{(s)}$.

\begin{figure*}[t]
    \includegraphics[width=2.\columnwidth]{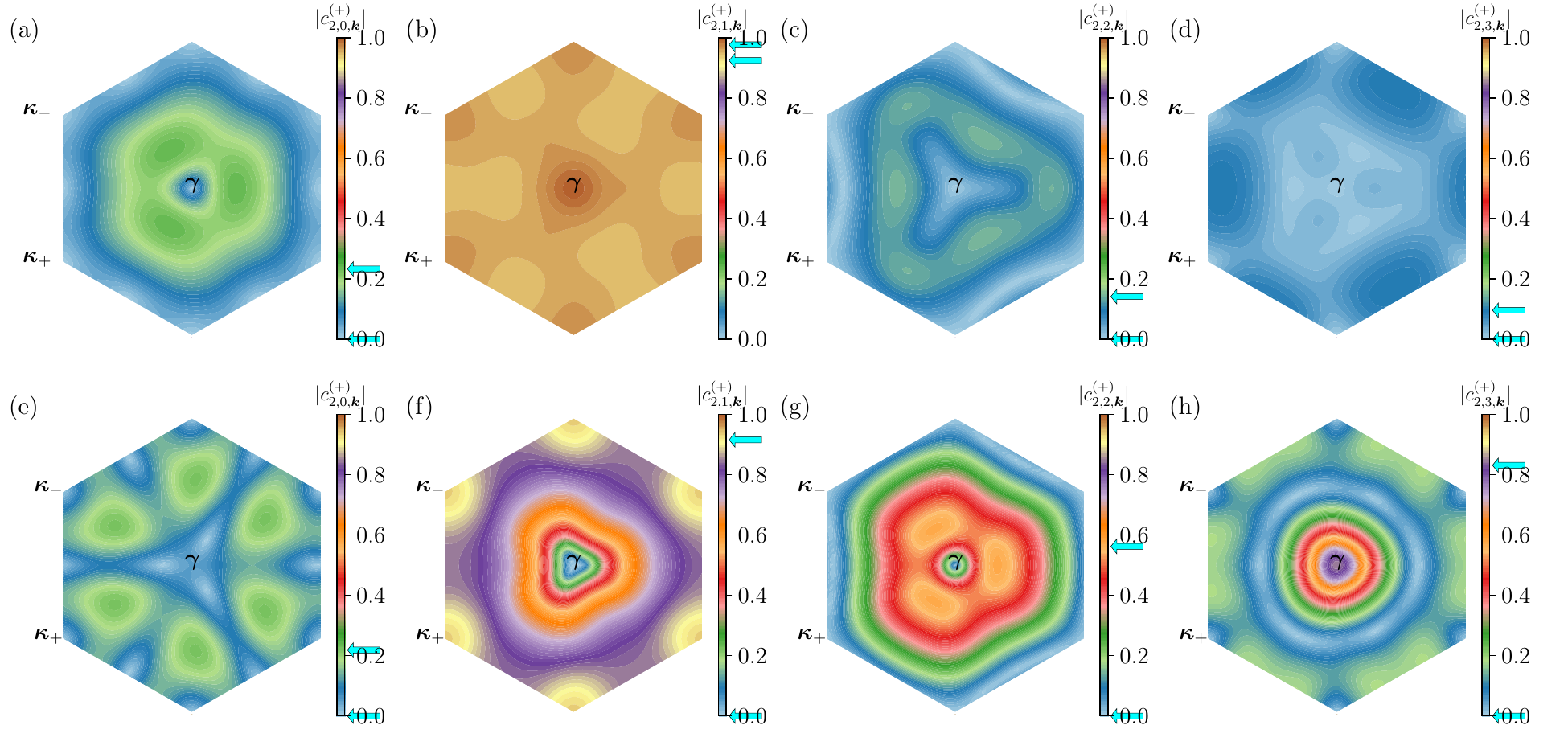}
    \caption{Overlap $\lvert c_{2,n,\bl{k}}^{(+)}\rvert=\lvert\langle\Theta^{(+)}_{n,\bl k}\lvert\psi_{+,2,\bl k}\rangle\rvert$ in mBZ for (a-d) $n=0,1,2,3$ at $\theta=2.13^\circ$ and (e-h) $n=0,1,2,3$ at $\theta=3.89^\circ$. The arrows indicate the data range in each plot.} 
    \label{fig:overlap_band2}
\end{figure*}

\subsection{The first moir\'e band} 
\label{section:IVA}

We first apply the variational mapping approach to the first moir\'e valence band in tMoTe$_2$. Based on its 0LL-like feature of quantum geometry with $\mathcal{C}=+1$ and $\mathcal{K} -|\mathcal{C}| \ll 1$, the dominant generalized LL weight is expected to be $W_{1,0}^{(+)}$. For our initial ansatz of $\mathcal{B}^{(+)}(\bl r)$, we take a simple ratio between the Bloch wave function 
$\psi_{+,1,\boldsymbol{\gamma}}(\bm{r})$ of the first moir\'e valence band at the 
$\boldsymbol{\gamma}$ point and the corresponding 0LL wave function 
$\Psi_{0,\boldsymbol{\gamma}}^{(+)}(\bm{r})$,
\begin{equation}
\label{Br0}
\begin{aligned}
\mathcal{B}^{(+)}(\bl r) \rightarrow
&\frac{\psi_{+,1,\bl\gamma}(\bl r)}{\Psi_{0,\bl\gamma}^{(+)}(\bl r)},
\end{aligned}    
\end{equation}
In this construction, any zeros in the numerator and denominator cancel, ensuring that this initial ansatz is continuous across the entire real space.  
After adjusting $\mathcal{B}^{(+)}(\bl r)$ according to Eq.~\eqref{iter} until convergence, its orthogonal counterpart $\mathcal{B}^{(-)}(\bm{r})$ is obtained via
\begin{equation}
\label{Br_relation}
\begin{aligned}
U^\dagger_0(\bl r)\mathcal{B}^{(-)}(\bl r)=i\sigma_y[U^\dagger_0(\bl r)\mathcal{B}^{(+)}(\bl r)]^*,
\end{aligned}    
\end{equation}
where $\sigma_y$ is the $y$ Pauli matrix in the layer pseudospin space. In addition to the orthogonality, this construction ensures that $\Theta^{(-)}_{n,\bl k}(\bl r)$ satisfies $C_{3z}$ symmetry at three high-symmetry points $\bl k\in\{\bl \kappa_\pm,\bl \gamma\}$, and respects $C_{2y}\mathcal{T}$ symmetry at $\bl \gamma$, as demonstrated in Appendix~\ref{appendix:A}.

The resulting generalized LL weights are shown in Figs.~\ref{fig:variation1}(a) and \ref{fig:variation1}(b).  
Throughout the entire $\theta$ range, $W_{1,0}^{(+)}$ is the dominant weight as anticipated and exceeds 0.9, while all other weights are smaller by at least an order of magnitude. The cumulative weight for $n \leq 4$, $\sum_{s=\pm}\sum_{n=0}^{4} W_{1,n}^{(s)}$, exceeds 0.99 for all twist angles studied.
$W_{1,0}^{(+)}$ attains its maximum value of 0.97 at $\theta = 2.45^\circ$, where $\mathcal{K} -|\mathcal{C}|$ is minimized. 
In Figs.~\ref{fig:variation1}(c-f) we present the map of  $\abs{\mathcal{B}^{(+)}(\bm{r})}$ in real space at four representative twist angles $\theta=2.13^\circ$, $2.45^\circ$, $2.88^\circ$, and $3.15^\circ$. Here $\abs{\mathcal{B}^{(+)}(\bm{r})}$ exhibits spatial modulations following the electron density variation of the first moir\'e band. At $\theta = 2.13^\circ$, $\lvert \mathcal{B}^{(+)}(\bm{r})\rvert$ displays relatively strong spatial fluctuation, with its maximal positions forming an effective triangular lattice. 
As $\theta$ increases, the arrangement of maxima in $\lvert \mathcal{B}^{(+)}(\bm{r})\rvert$ gradually evolves from the triangular to its dual honeycomb geometry.
Meanwhile, the spatial fluctuations in $\lvert \mathcal{B}^{(+)}(\bm{r})\rvert$ weaken as $\theta$ increases from $2.13^{\circ}$ to $2.88^{\circ}$, but re-intensifies at larger $\theta$. 

Figure~\ref{fig:overlap_band1} shows $\lvert c_{1,n,\bm{k}}^{(s)} \rvert$ in momentum space for $s = \pm$ and $n = 0,1$ at $\theta = 2.45^\circ$. Each amplitude $\lvert c_{1,n,\bm{k}}^{(s)} \rvert$ exhibits threefold rotational symmetry in the mBZ and is invariant under $k_y \rightarrow -k_y$, as a consequence of the $C_{3z}$ and $C_{2y}\mathcal{T}$ symmetries. Notably, the dominant coefficient $\lvert c_{1,0,\bm{k}}^{(+)} \rvert$ is above 0.98 throughout the mBZ, and it maximizes at the $\boldsymbol{\gamma}$ point, reaching nearly 1.0, thereby confirming the validity of the ansatz in Eq.~\eqref{Br0}. An adiabatic connection between the moir\'e band wave function $\psi_{+,1,\bm{k}}(\bl r)$ and the generalized $0$LL can be established by gradually tuning $c^{(s)}_{1,n,\bm{k}}$ to the configuration where $c_{1,0,\bm{k}}^{(+)} = 1$ and all other coefficients vanish. We further note that $c_{1,1,\bm{k}}^{(+)}$ vanishes at the ${C}_{3z}$ invariant momenta $\boldsymbol{\gamma}$, $\bm{\kappa}_+$, and $\bm{\kappa}_-$, because the wave functions  $\psi_{+,1,\bm{k}}(\bl r)$ and $\Theta^{(+)}_{1,\bl k}$ carry different angular momenta under ${C}_{3z}$ at these momenta and are therefore orthogonal. The same reasoning explains the zeros of $c_{1,0,\bm{k}}^{(-)}$ at $\boldsymbol{\gamma}$ and of $c_{1,1,\bm{k}}^{(-)}$ at $\boldsymbol{\gamma}$ and $\bm{\kappa}_\pm$.

\subsection{The second moir\'e band} 
We turn to the second moir\'e valence band of $t$MoTe$_2$. Near $\theta = 2.13^{\circ}$, this band exhibits quantum-geometric features analogous to the generalized 1LL, characterized by $\mathcal{C}=+1$ and $\mathcal{K}=3.36$. Since $\mathcal{K}$ is close to $3\mathcal{C}$, we choose the dominant generalized LL weight to be $W_{2,1}^{(+)}$. The Chern number of this band changes sign for $\theta \geq 3.15^{\circ}$ due to band inversion with the third band at the $\boldsymbol{\gamma}$ point. Since the associated wave-function changes are mainly for momenta around $\boldsymbol{\gamma}$, $W_{2,1}^{(+)}$ should remain as the leading weight. We therefore maximize $W_{2,1}^{(+)}$ across the full range of $\theta$ considered.  For the initial ansatz, we use $\mathcal{B}^{(+)}(\bm{r})$ obtained from optimizing $W_{1,0}^{(+)}$, and refine it to enhance $W_{2,1}^{(+)}$ within the variational scheme described above. Finally, $\mathcal{B}^{(-)}(\bm{r})$ is obtained again from Eq.~\eqref{Br_relation}.

Figures~\ref{fig:band2}(a) and \ref{fig:band2}(b) present the resulting generalized LL weights for the second moir\'e valence band.  In this case, $W_{2,1}^{(+)}$ indeed dominates for $\theta \in [2.13^\circ, 3.89^\circ]$, reaching a maximum of 0.91 at $\theta = 2.13^\circ$ and decreasing to 0.6 at $\theta = 3.89^\circ$.  The second-largest weight $W_{2,2}^{(+)}$ rises to 0.1 at $\theta = 3.89^\circ$, while all $W_{2,n}^{(-)}$ remain below 0.1. Importantly, the emergence of a dominant generalized $1$LL component is nontrivial: although the quantum weight $\mathcal{K}$ of the second band lies close to $3|\mathcal{C}|$, this condition alone does not a priori ensure a generalized $1$LL–like wave function. Therefore, the explicit decomposition into generalized LLs is essential for unambiguously establishing the $1$LL-like nature of the second moir\'e band.

The spatial profiles of the resulting $|\mathcal{B}^{(+)}(\bl{r})|$ are shown in Figs.~\ref{fig:band2}(c) and \ref{fig:band2}(d) for $\theta = 2.13^\circ$ and $\theta = 2.45^\circ$, respectively, where the arrangement of maxima evolves from a triangular to its dual honeycomb geometry, similar to the first moir\'e band.
The $\theta$ dependence of the generalized LL weights is further reflected in momentum space, as illustrated in Fig.~\ref{fig:overlap_band2}. This figure displays $|c_{2,n,\bm{k}}^{(+)}|$ for $n=0$–$3$ at $\theta = 2.13^\circ$ and $3.89^\circ$, corresponding to Chern numbers $\mathcal{C} = +1$ and $-1$, respectively, for the second moir\'e band. These amplitudes share the same symmetries as $|c_{1,n,\bm{k}}^{(s)}|$ in Fig.~\ref{fig:overlap_band1}, including threefold rotational symmetry and invariance under $k_y \to -k_y$. At $\theta = 2.13^\circ$, the dominant component $|c_{2,1,\bm{k}}^{(+)}|$ is above 0.94 across the mBZ, indicating an adiabatic connection between $\psi_{+,2,\bm{k}}(\bl r)$ and the generalized 1LL. In contrast, no single component dominates at $\theta = 3.89^\circ$ for every momentum, confirming that $\psi_{+,2,\bm{k}}(\bl r)$ becomes topologically distinct. In particular, at the $\boldsymbol{\gamma}$ point, $|c_{2,3,\bm{k}}^{(+)}|$ is dominant and reaches 0.9, whereas $|c_{2,1,\bm{k}}^{(+)}|$ vanishes; away from $\boldsymbol{\gamma}$, $|c_{2,1,\bm{k}}^{(+)}|$ becomes dominant, approaching unity at $\boldsymbol{\kappa}_{\pm}$.

\subsection{Remarks}

We make two remarks about the decomposition. First, in the variational decomposition, we first select a candidate generalized LL basis guided by physical inputs (e.g., quantum geometry, Chern number, quantum weight, and symmetry eigenvalues), and then optimizes $\mathcal{B}(\bm r)$ to maximize the overlap with that chosen basis. The method can thus serve as a diagnostic tool for quantifying how well a physically motivated generalized LL description captures the underlying band structure. When the dominant LL character is not known a priori, the method can be adapted to systematically compare multiple candidate generalized LL bases and evaluate the resulting optimized overlaps. The relative quality of these decompositions can then be used to identify the most appropriate emergent LL description.

Second, to demonstrate the robustness of our variational method, we repeat the mapping using a random complex function as the initial $\mathcal{B}^{(+)}(\bl{r})$ and compare the results with those obtained from the physically motivated ansatz used in Figs. \ref{fig:variation1} and \ref{fig:band2}. The two procedures yield essentially indistinguishable optimized $\mathcal{B}(\bl{r})$ across the twist angles. The resulting generalized LL weights agree quantitatively, confirming that the optimization converges to a unique solution independent of the initial ansatz.

\section{Abelian Fractionalized States} 
\label{sec:Abelian_FCI}

\begin{figure}[b]
\includegraphics[width=1.\columnwidth]{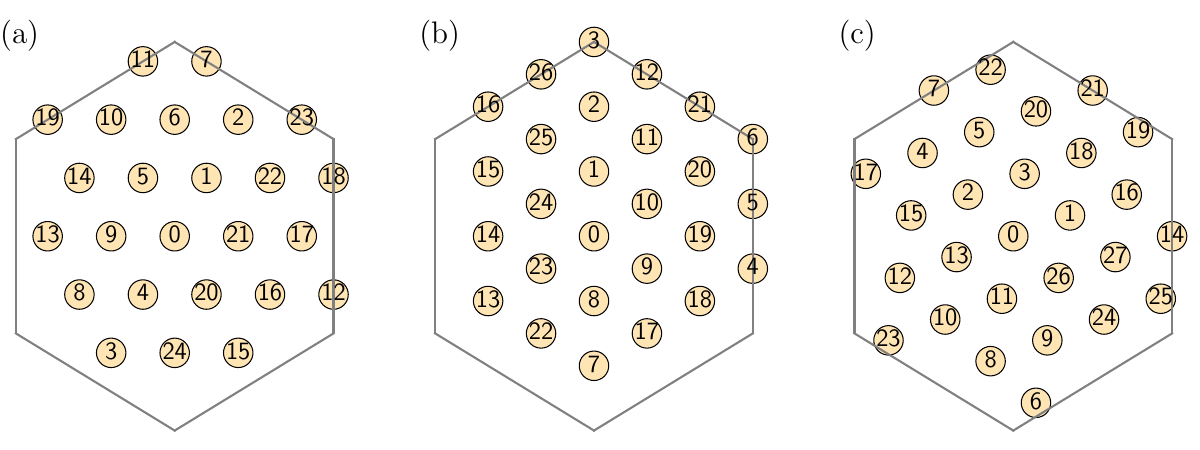}
    \caption{(a-b) The 25-, 27-, and 28-unit-cell momentum clusters used in the ED calculations. The circled numbers are momentum indices. } 
    \label{fig:cluster1}
\end{figure}

We investigate the many-body effects within the first moir\'e band. Given its close similarity to the generalized 0LL, Abelian fractionalized states can be expected. The many-body Hamiltonian projected onto the first band is given by,
\begin{align}
\label{H1}
\hat{\mathcal{H}}_1=\mathcal{P}_1 \hat H^{\mathrm{(full)}} \mathcal{P}_1.
\end{align}
where $\hat H^{\mathrm{(full)}}$ is the full many-body Hamiltonian in the hole basis (see Appendix \ref{appendix:C} for details). Here the subspace $\mathcal{P}_1$ is spanned by the many-body basis states where the first band at the $+K$ valley is partially occupied and all other bands are empty in the hole basis,
\begin{align}
\label{subspace}
\mathcal P_1=\mathrm{span}\{\prod_{i=1}^{N_e}\varphi^\dagger_{+,1,\bl {k}_i}\ket{0},\bl k_i\in\mathrm{mBZ}\},
\end{align}
where $\varphi_{\tau,n,\bl k}^{\dagger}$ ($\varphi_{\tau,n,\bl k}$) denotes the hole creation (annihilation) operator of a Bloch state at momentum $\bl k$, $n$th band and valley $\tau$, $N_e$ is the number of doped holes and $\ket{0}$ is the vacuum state corresponding to no doped holes in tMoTe$_2$. We assume full valley polarization for the doped holes in the projected Hamiltonian $\hat{\mathcal{H}}_1$, which can be further formulated as,
\begin{equation}
\begin{aligned}
&\hat{\mathcal{H}}_1= \sum_{\bl{k}}(-\mathcal{E}_{+,1,\bl{k}})\varphi^\dagger_{+,1,\bl{k}}\varphi_{+,1,\bl{k}}\\
&+\sum_{\bl{k_1k_2k_3k_4}}V_{\bl{k_1k_2k_3k_4}}
\varphi^\dagger_{+,1,\bl{k_1}}\varphi^\dagger_{+,1,\bl{k_2}}
\varphi_{+,1,\bl{k_3}}\varphi_{+,1,\bl{k_4}},
\end{aligned}   
\label{Hproj}
\end{equation}
The interaction matrix element $V_{\bl{k_1k_2k_3k_4}}$ is given by
\begin{equation}
\begin{aligned}
    V_{\bl{k_1k_2k_3k_4}}=&\frac{1}{2\mathcal{A}}\sum_{\bl{q}}V(\bl q)
    M^{+ 11}_{\bl{k_1}\bl{k_4}}(\bl{q})
    M^{+ 11}_{\bl{k_2}\bl{k_3}}(\bl{-q}),\\
\end{aligned}
\label{IntV1}
\end{equation}
where $\mathcal{A}$ is the system area, $V_{\bl{q}}=2\pi e^2/(\epsilon\lvert\bl{q}\rvert)$ is the Coulomb interaction with $\epsilon$ the dielectric constant, and we set $\epsilon=5$ in this work. The plane-wave matrix element $M_{\bl{k}\bl{k'}}^{\tau nn'}(\bl{q})$ is defined as
\begin{equation}
\begin{aligned}
M_{\bl{k}\bl{k'}}^{\tau nn'}(\bl{q})
=&\int d\bl{r}\,e^{i\bl{q}\cdot \bl{r}}[f_{\tau,n,\bl{k}}(\bl{r})]^*f_{\tau,n',\bl{k}'}(\bl{r}),
\end{aligned}
\label{inner}
\end{equation}
where $f_{\tau,n,\bl{k}}(\bl{r})$ denotes the Bloch wave function. In Eq.~\eqref{inner}, we take $f_{+,1,\bl{k}}(\bl{r})$ to be $[\psi_{+,1,\bl k}(\bl r)]^*$ and $[\Theta_{0,\bl{k}}^{(+)}(\bl r)]^*$, which we refer to as the original model and the variational model, respectively. Here $\psi_{+,1,\bl k}(\bl r)$ is the Bloch wave function for the first moir\'e band of the single particle Hamiltonian $H_+$ in Eq.~\eqref{H0}, and $\Theta_{0,\bl{k}}^{(+)}(\bl r)$ is the generalized 0LL wave function variationally obtained in Section~\ref{section:IVA}. The complex conjugation is taken to implement the particle-hole transformation, as we work in the hole basis.

\begin{figure}[t]    \includegraphics[width=1.\columnwidth]{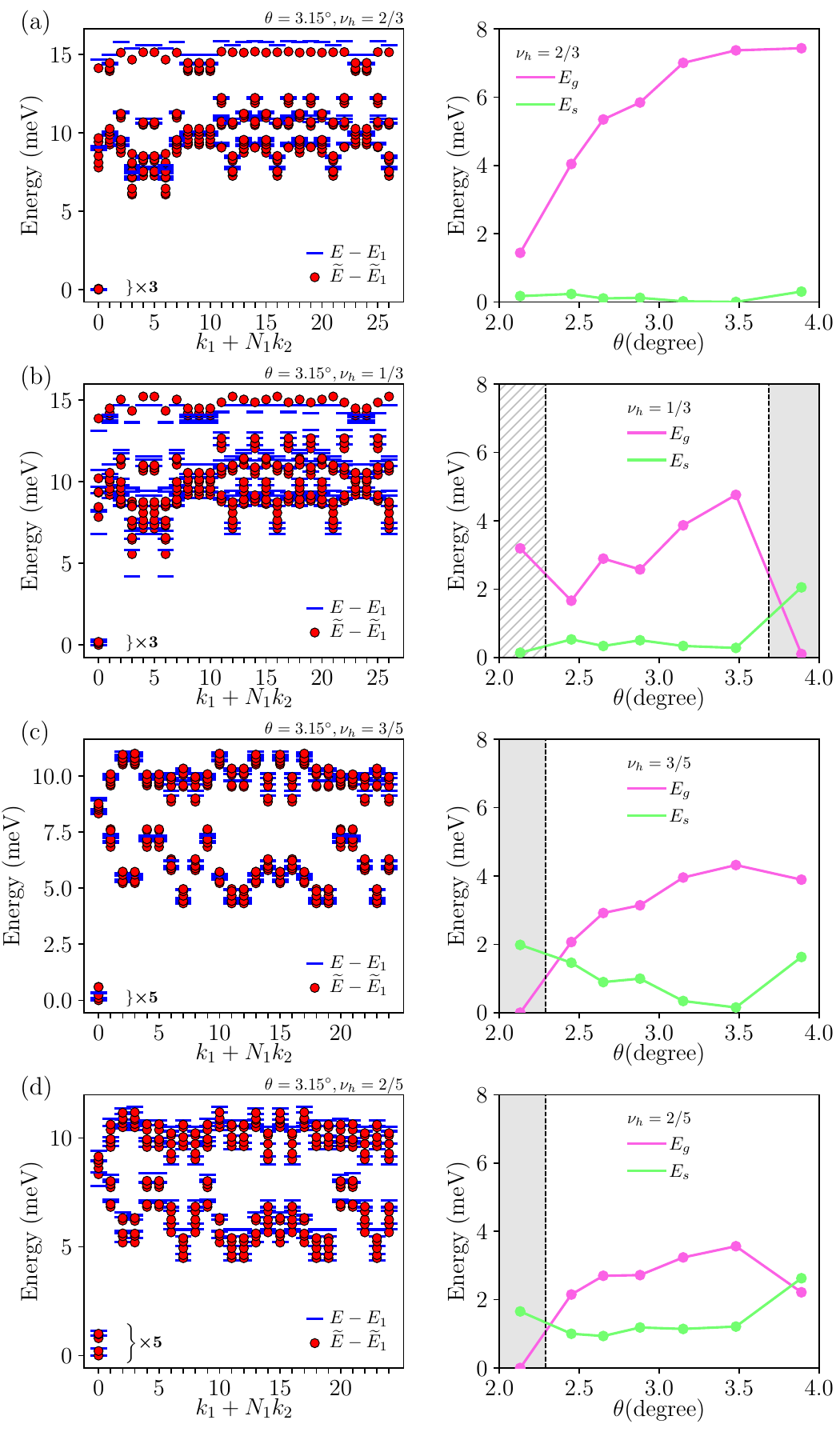}
    \caption{(a-d) ED results at $\nu_h=2/3,1/3,3/5$, and $2/5$. Left panels, ED spectrum of the original model (blue lines) and variational model (red dots) at $\theta=3.15^\circ$. Right panels,  The energy gap $E_g$ (pink) and spread $E_s$ (green) as functions of $\theta$ based on the original model. Gray regions mark the gapless phase, while the hatched region denotes the CDW phase.} 
    \label{fig:ED}
\end{figure}

\begin{figure}[t] \includegraphics[width=1.\columnwidth]{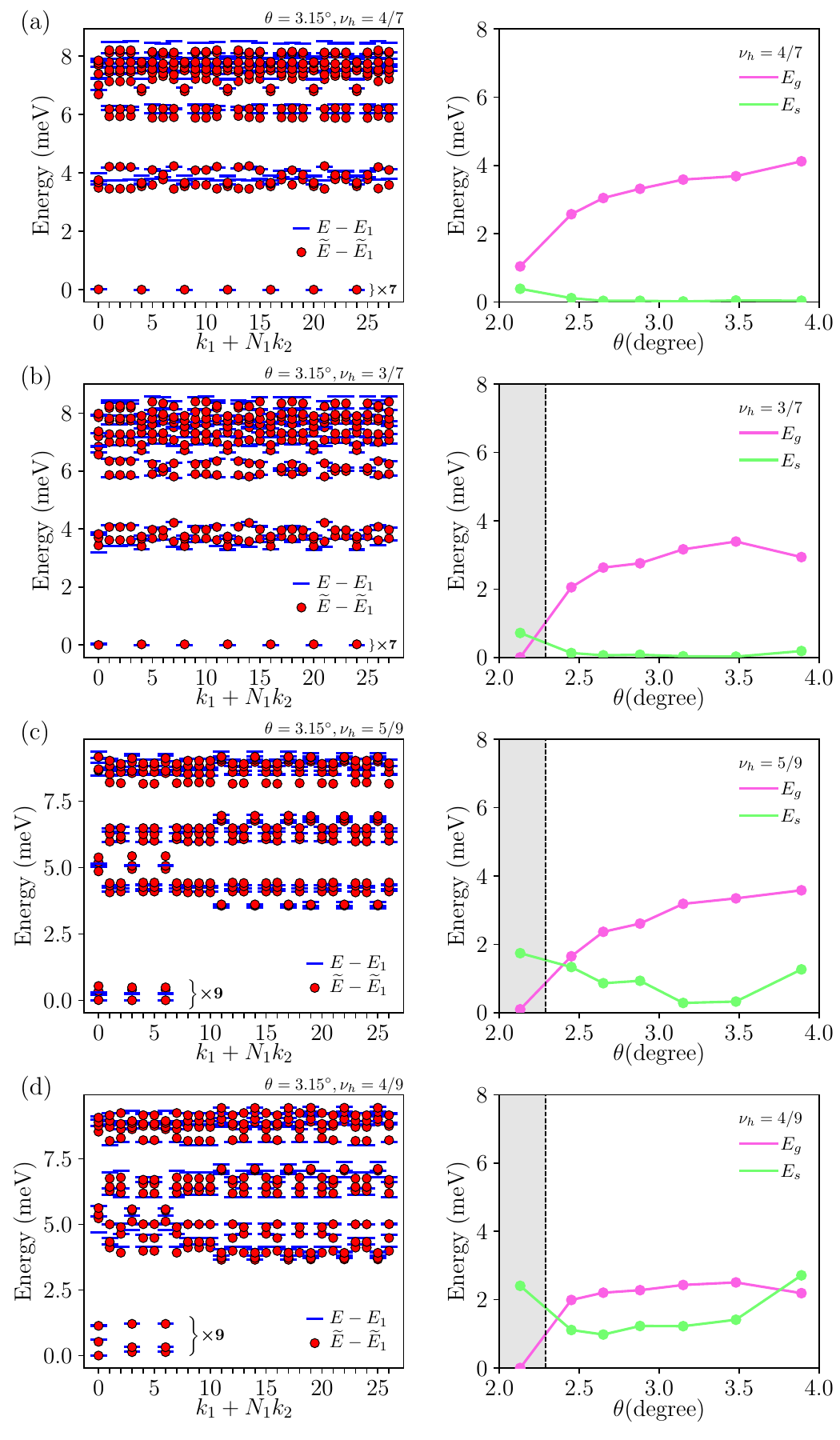}
    \caption{(a-b) ED results at $\nu_h=4/7,3/7,5/9$, and $4/9$. Left panels, ED spectrum of the original model (blue lines) and variational model (red dots) at $\theta=3.15^\circ$. Right panels,  The energy gap $E_g$ (pink) and spread $E_s$ (green) as functions of $\theta$ based on the original model. Gray regions mark the gapless phase.} 
    \label{fig:ED2}
\end{figure}

We perform ED for the projected Hamiltonian $\hat{\mathcal{H}}_1$. Our study focuses on hole fillings $\nu_h = 1/3$ and $2/5$, which belong to the Jain sequence $\nu_h = p/(2p+1)$ with $p = 1, 2$, as well as their particle-hole conjugates $\nu_h = 2/3$ and $3/5$. The ED calculations are carried out on a 27-unit-cell cluster for $\nu_h = 1/3$ and $2/3$, and on a 25-unit-cell cluster for $\nu_h = 2/5$ and $3/5$. 
In Figs.~\ref{fig:cluster1}(a) and \ref{fig:cluster1}(b) we show the geometries of these clusters in the momentum space. In Fig.~\ref{fig:ED}, the left panels show the ED spectra for both models at $\theta = 3.15^\circ$, which exhibit quantitative agreement and clearly reveal the characteristics of Abelian-type FCIs at the Jain sequences.
Specifically, the threefold quasi-degenerate ground states appear at zero momentum for $\nu_h = 1/3$ and $2/3$, and fivefold quasi-degenerate ground states appear at zero momentum for $\nu_h = 2/5$ and $3/5$, which are separated from excited states by a finite gap. 

\begin{figure}[t]
    \includegraphics[width=1.\columnwidth]{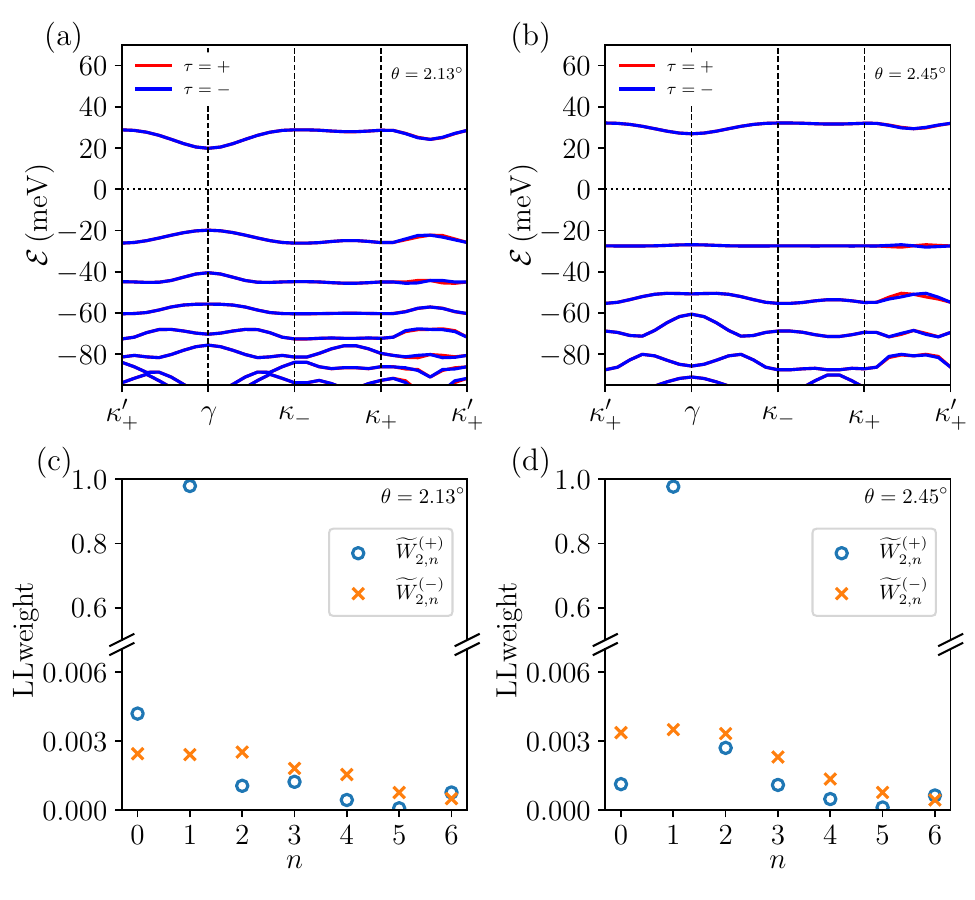}
    \caption{(a-b) HF band structure presented in the electron basis. Filling factor $\nu_h$ is 2 and twist angle $\theta$ is $2.13^{\circ}$ and $2.45^{\circ}$. The bands in $\tau K$ valley are plotted by red and blue lines for $\tau=+$ and $-$, respectively. The horizontal dotted line mark the Fermi energy in the gap. (c-d) LL weights $\wt{W}_{2,n}^{(s)}$ at $2.13^{\circ}$ and $2.45^{\circ}$.} 
    \label{fig:nonabelian}
\end{figure}

 We present the twist angle dependence of the charge-neutral gap $E_g$ and ground state energy spread $E_s$ in the right panels of Fig.~\ref{fig:ED}. We define the charge-neutral gap as $E_g = E_{n+1} - E_n$ and the ground-state energy spread as $E_s = E_n - E_1$, where $E_i$ is the $i$-th lowest energy level. We set $n = 3$ for $\nu_h = 1/3$ and $2/3$, and  $n = 5$ for $\nu_h = 2/5$ and $3/5$. 
 The numerical results can be summarized as follows.
 (1) At $\nu_h=2/3$, $E_g$ is significantly larger than $E_s$ within $\theta\in[2.13^\circ,3.89^\circ]$, suggesting a stable FCI. Furthermore, $E_g$ grows with increasing $\theta$, which can be attributed to the enhancement of the characteristic interaction energy scale $e^2/(\epsilon a_M)$.
(2) At $\nu_h=1/3$, the FCI phase is stable only in a narrower twist-angle window, $\theta\in[2.45^\circ,3.48^\circ]$, as characterized by a finite $E_g$. Beyond this region, the gap closes at $\theta = 3.89^\circ$, while the system transitions into a CDW state at $\theta = 2.13^\circ$. The CDW phase is characterized by a threefold quasi-degenerate ground state across momentum sectors at the mBZ center and corners, a transition driven by enhanced spatial variation of the single-particle wave functions. 
(3) At $\nu_h=3/5$ and $2/5$, the FCI phase remains stable within $\theta\in[2.45^\circ,3.89^\circ]$, as evidenced by a finite energy gap $E_g$. The system becomes gapless below this range at $\theta=2.13^\circ$. 

We also examine fillings $\nu_h = 3/7$ and $4/9$, which belong to the Jain sequence $\nu_h = p/(2p+1)$ with $p = 3, 4$, as well as their particle-hole conjugates $\nu_h = 4/7$ and $5/9$. The ED calculations are carried out on a 28-unit-cell cluster for $\nu_h = 3/7$ and $4/7$, and on a 27-unit-cell cluster for $\nu_h = 4/9$ and $5/9$, with momentum clusters shown in Fig.~\ref{fig:cluster1}. In Fig.~\ref{fig:ED2}, the left panels show the ED spectra for the original and variational models at $\theta = 3.15^\circ$, which exhibit quantitative agreement and clearly reveal the characteristics of Abelian-type FCIs at the Jain sequences.
These features are evidenced by the sevenfold quasi-degenerate ground states appear for $\nu_h = 3/7$ and $4/7$, and ninefold quasi-degenerate ground states appear for $\nu_h = 4/9$ and $5/9$, which are separated from excited states by a finite gap. 

We show $E_g$ and $E_s$ as a function of $\theta$ in the right panels of Fig.~\ref{fig:ED2}, where we set $n = 7$ for $\nu_h = 3/7$ and $4/7$, and $n = 9$ for $\nu_h = 4/9$ and $5/9$. At $\nu_h=4/7$, $E_g$ is significantly larger than $E_s$ within $\theta\in[2.13^\circ,3.89^\circ]$, suggesting a stable FCI. At $\nu_h=3/7$ and $5/9$, the FCI phase remains stable with a finite energy gap $E_g$ for $\theta\in[2.45^\circ,3.89^\circ]$, and the system becomes gapless below this range at $\theta=2.13^\circ$. At $\nu_h = 4/9$, $E_g$ is finite but comparable to $E_s$ for $\theta \in [2.45^\circ, 3.89^\circ]$, indicating that the FCI phase is less robust. Additionally, we perform ED calculations at $\nu_h = 1/2$. The ED spectra at $\nu_h = 1/2$ mimics that of the 0LL for all twist angles under study, indicating a composite Fermi liquid state (see Appendix \ref{appendix:D} for details).

The asymmetry between the FCI states at $\nu_h = p/(2p+1)$ and $(p+1)/(2p+1)$ arises from the lack of particle–hole symmetry in the first band of tMoTe$_2$, stemming from its finite bandwidth and momentum-dependent quantum geometric tensor, in contrast to conventional LLs. As a general trend, we find that FCI states in the range $1/2 < \nu_h < 1$ are more robust than their particle–hole conjugates at $\nu_h < 1/2$, as evidenced by their larger gaps. This asymmetry reflects a subtle interplay between the single-particle wave function, bandwidth and electron interactions, which invites further exploration.

\begin{figure}[t]
 \includegraphics[width=1.\columnwidth]{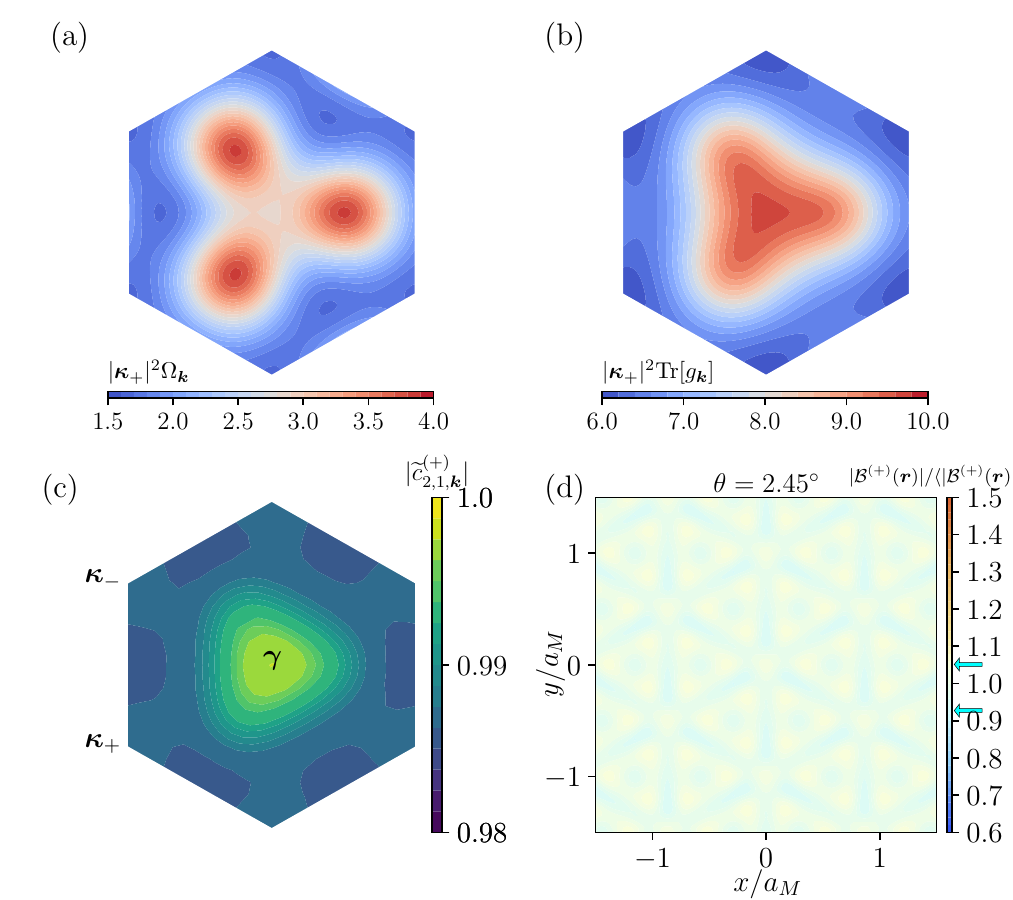}
    \caption{(a-b) Berry curvature $\Omega_{\bl k}$ and trace of quantum metric $\mathrm{Tr}[g_{\bl k}]$ in the mBZ for  the second HF band in $+K$ valley. 
    (c) Overlap $\lvert\wt{c}_{2,1,\bl k}^{(+)}\rvert$.
    (d) Map of $\lvert \mathcal B^{(+)}(\bl r)\rvert$ scaled by its spatial average. In all plots, $\theta=2.45^\circ$.  } 
    \label{fig:nonabelian2}
\end{figure}

\section{nonabelian fractionalized state}
\label{sec:nonAbelian_FCI}
The close similarity between the second band of tMoTe$_2$ and the generalized 1LL at $\theta = 2.13^\circ$ suggests that tMoTe$_2$ may provide a viable platform for realizing a non-Abelian phase. Motivated by this connection, we focus on $\nu_h = 5/2$, a candidate filling factor for stabilizing the MR state. A multi-band ED analysis at this filling, however, is computationally prohibitive. To address this challenge, we first carry out a self-consistent HF calculation at $\nu_h = 2$, which allows us to construct a refined Bloch basis renormalized by interactions.

Using this refined basis, we then build the effective many-body Hamiltonian at $\nu_h = 5/2$ by treating the first renormalized band in both valleys as inert and considering a half-filled second band in the $+K$ valley, assuming spontaneous valley polarization. This Hamiltonian is subsequently studied using ED, and the resulting energy spectrum and PES are analyzed to assess whether the system exhibits features consistent with the MR phase.

\subsection{HF Renormalized Bands} 
\label{section:VIA}
We first perform HF calculation at $\nu_h = 2$, approximating the interaction effects via corrections to the single-particle orbitals (described in Appendix \ref{appendix:C}). This yields a corrected Bloch wave function, denoted as $[\wt \psi_{\tau,n,\bl k}(\bl r)]^*$ in the hole basis. The corresponding creation and annihilation operators are $\wt\varphi^\dagger_{\tau,n,\bl k}$ and $\wt\varphi_{\tau,n,\bl k}$.

\begin{figure}[b]   \includegraphics[width=\columnwidth]{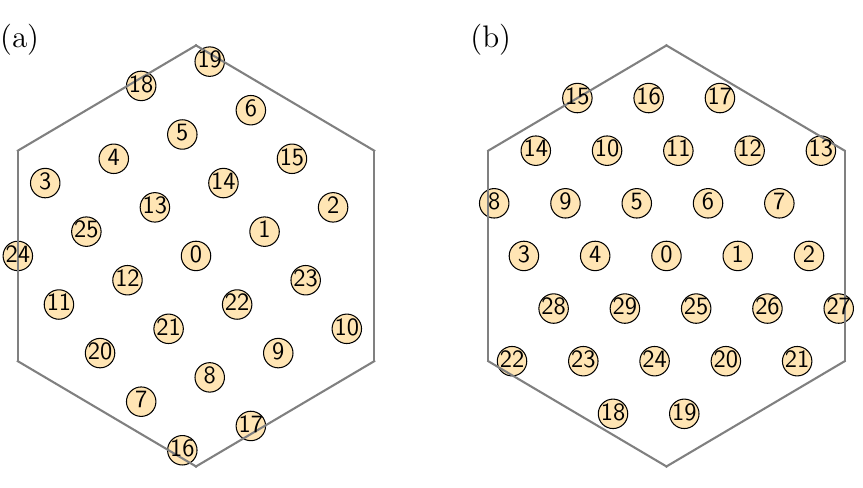}
    \caption{(a-b) The 26- and 30-unit-cell momentum clusters used in the ED calculations. The circled numbers are momentum indices. }
    \label{fig:nonabelian-ED-clusters}
\end{figure}

\begin{figure*}[t]
    \includegraphics[width=\textwidth]{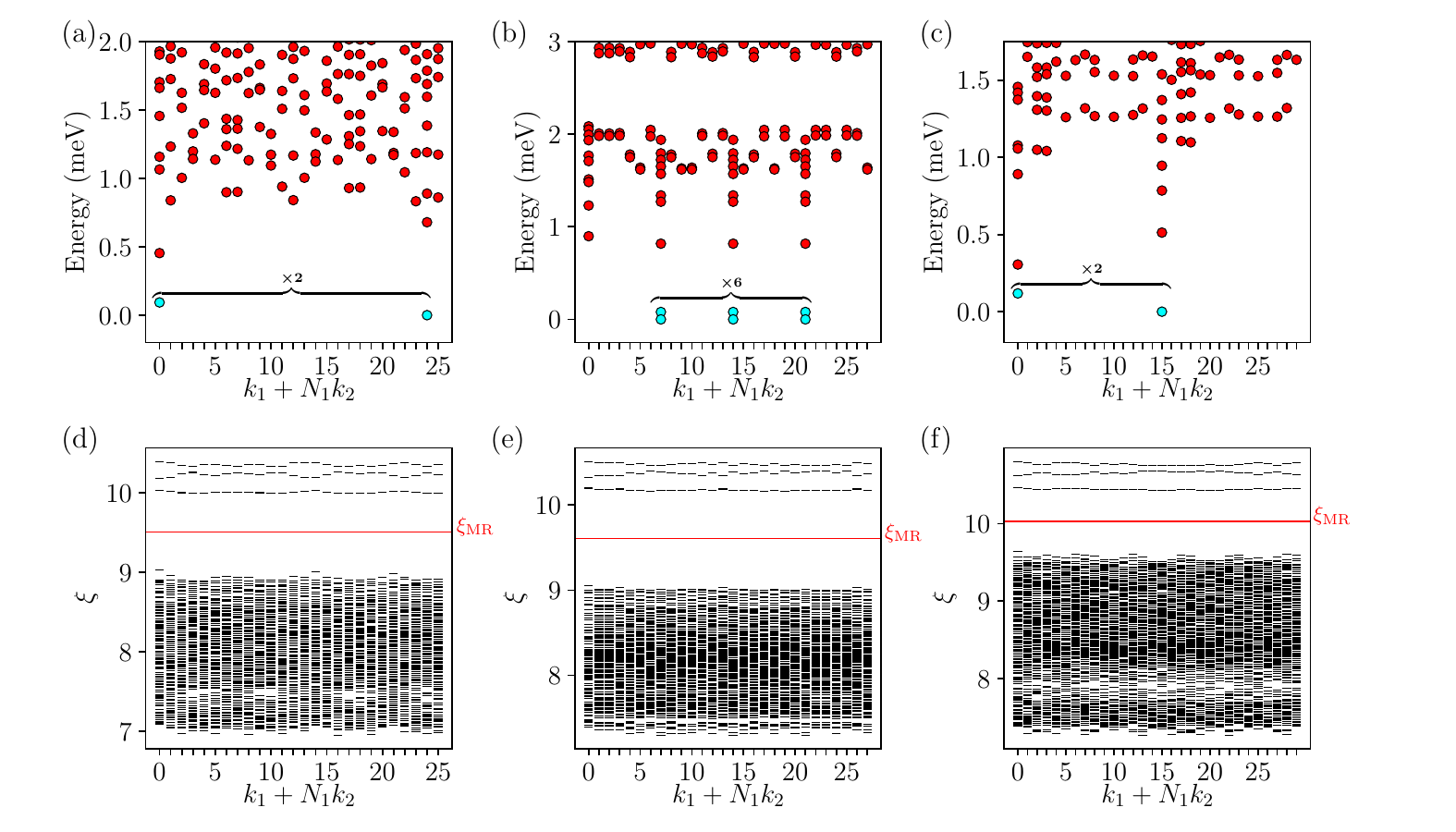}
    \caption{(a-c) ED spectrum at $\nu_h=5/2$ and $\theta = 2.45^\circ$, obtained with $N=26,28$, and 30 clusters. The quasi-degenerate states are highlighted in cyan. (d-f) PES with $N_A=3$ for the quasi-degenerate states highlighted in cyan in (a-c). }
    \label{fig:nonabelian-ED1}
\end{figure*}

Figures \ref{fig:nonabelian}(a) and \ref{fig:nonabelian}(b) show the HF band structure at $\nu_h = 2$ with twist angle $\theta = 2.13^\circ$ and $\theta = 2.45^\circ$, respectively. In both cases, the top four moir\'e bands in the electron basis are isolated from each other and exhibit narrow bandwidths. All of these bands possess a quantized Chern number $\mathcal{C}=+1$ in the $+K$ valley. 
The quantum weight $\mathcal{K}$ of the $n$th topmost band for $n = 1$ to $4$ is $1.03$, $3.08$, $5.17$, and $7.37$ at $\theta = 2.13^\circ$, and $1.02$, $3.08$, $5.18$, and $7.45$ at $\theta = 2.45^\circ$, respectively.
The deviation between the $n$th band and the generalized $(n-1)$LL Landau level, quantified by $\mathcal{K} - (2n-1)|\mathcal{C}|$, remains  below 0.5 for all four bands. This suggests a series of LL-like bands after the HF correction.

Focusing on the second band,  the corresponding quantum geometric quantities at $\theta=2.45^\circ$, namely, $\Omega_{\bl{k}}$ and $\mathrm{Tr}[ g_{\bl{k}}]$, are displayed, respectively, in Figs.~\ref{fig:nonabelian2}(a) and \ref{fig:nonabelian2}(b), both exhibiting strong momentum dependence. Nevertheless, the indicator $\mathcal{K}-3|\mathcal{C}|$ takes a small value of $0.08$, reduced compared to its noninteracting value of $0.50$ . This reduction has also been observed in Ref.~\onlinecite{Wang2025Higher}. The renormalized bandwidth is 1.3 meV at $\theta = 2.45^\circ$, which is smaller than the noninteracting value, and increases to 6.3 meV at $\theta = 2.13^\circ$.

We further perform the variational mapping between the Bloch wave function $\wt{\psi}_{+,2,\bl k}(\bl r)$ and the generalized 1LL. 
The dominant weight is chosen as $\wt W_{2,1}^{(+)}$, where we define
\begin{align}
\wt W_{m,n}^{(s)}=\frac{1}{N}\sum_{\bl k}\lvert  \wt c_{m,n,\bl{k}}^{\,(s)}\rvert^2,
\end{align}
and $\wt c_{m,n,\bl{k}}^{\,(s)}=\langle\Theta^{(s)}_{n,\bl k}\lvert\wt \psi_{+,m,\bl k}\rangle$. 
Following the procedure described in Section~\ref{sec:mapping}, we variationally adjust $\mathcal{B}^{(+)}(\bl r)$ in $\Theta^{(+)}_{n,\bl k}(\bl r)$ to maximize $\wt W_{2,1}^{(+)}$.
Figures~\ref{fig:nonabelian}(c) and \ref{fig:nonabelian}(d) present the resulting generalized LL weight $\wt W_{2,n}^{(s)}$ of the second band at $\theta = 2.13^\circ$ and $\theta = 2.45^\circ$, respectively. The dominant weight $\wt W_{2,1}^{(+)}$ reaches 0.97 at both twist angles—larger than its noninteracting value—while all other weights are smaller by two orders of magnitude. At $\theta = 2.45^\circ$, the overlap between $\wt{\psi}_{+,2,\bl k}(\bl r)$ and the generalized 1LL wave function, $\lvert\wt{c}_{2,1,\bl k}^{(+)}\rvert$, exceeds 0.98 across the entire mBZ, as shown in Fig.~\ref{fig:nonabelian2}(c). We also display the map of $\lvert \mathcal B^{(+)}(\bl r)\rvert$ in Fig.~\ref{fig:nonabelian2}(d), which exhibits suppressed fluctuations relative to the generalized 1LL calculated from the noninteracting wave function ${\psi}_{+,2,\bl k}(\bl r)$, as shown in Fig.~\ref{fig:band2}(d).

\subsection{ED Calculation} 
\label{section:VIB}

We now carry out ED studies at $\nu_h=5/2$ in the projected many-body Hamiltonian,
\begin{align}
\label{H2}
\hat{\mathcal{H}}_2=\mathcal{P}_2 \hat H^{\mathrm{(full)}} \mathcal{P}_2.
\end{align}
Here $\mathcal{P}_2$ is spanned by the many-body basis states where the first band is completely filled and the second band is partially occupied at $+K$ valley,
\begin{align}
\label{subspace}
\mathcal P_2=\mathrm{span}\{\prod_{i=1}^{N_e'}\wt\varphi^\dagger_{+,2,\bl {k}_i}\prod_{\tau=\pm}\prod_{j=1}^{N}\wt\varphi^\dagger_{\tau,1,\bl {k}_j}\ket{0},\bl k_i\in\mathrm{mBZ}\},
\end{align}
$N$ is the number of momentum points (unit cells) in the momentum (real) space, $N_e'=N/2$ is the number of holes in the second band, and $N_e'+2N$ is the totoal number of holes.
$\hat{\mathcal{H}}_2$ is formulated as,

\begin{align}
\hat{\mathcal{H}}_2 = \sum_{\bl k}(-\lambda_w \wt{\mathcal E}_{+,2,\bl k})\wt{\varphi}^\dagger_{+,2,\bl k}\wt{\varphi}_{+,2,\bl k}+\hat V+E_1,
\label{model_nonabelian}
\end{align}
where $-\wt{\mathcal{E}}_{\tau,n,\bl k}$ is the HF corrected band energy in the hole basis and $E_{1}$ denotes the total energy of the filled first band. We introduce a parameter $\lambda_w$ to phenomenologically tune the bandwidth. In $\hat{\mathcal{H}}_2$, $\hat V$ is the interaction term projected onto the second band,
\begin{equation}
\begin{aligned}
&\hat V=\sum_{\bl{k_1k_2k_3k_4}}\wt{V}_{\bl{k_1k_2k_3k_4}}
\wt{\varphi}^\dagger_{+,2,\bl{k_1}}\wt{\varphi}^\dagger_{+,2,\bl{k_2}}
\wt{\varphi}_{+,2,\bl{k_3}}\wt{\varphi}_{+,2,\bl{k_4}}.
\end{aligned}
\end{equation} 
The interaction matrix element $\wt{V}_{\bl k_1\bl k_2\bl k_3\bl k_4}$ is given by
\begin{equation}
\begin{aligned}
\wt{V}=&\frac{1}{2\mathcal{A}}\sum_{\bl{q}}V(\bl q)
\wt M^{+ 22}_{\bl{k}_1\bl{k}_4}(\bl{q})
\wt M^{+22}_{\bl{k}_2\bl{k}_3}(\bl{-q}),
\end{aligned}
\label{intV_opt}
\end{equation}
and the plane-wave matrix element is 
\begin{equation}
\begin{aligned}
\wt M_{\bl{k}\bl{k'}}^{+22}(\bl{q})
=&\int d\bl{r}\,e^{i\bl{q}\cdot \bl{r}}[\wt f_{+,2,\bl{k}}(\bl{r})]^*\wt f_{+,2,\bl{k}'}(\bl{r}),
\end{aligned}
\label{Mkk}
\end{equation}
and we take the wave function $\wt f_{+,2,\bl{k}}(\bl{r})$ to be $[\wt{\psi}_{+,2,\bl k}(\bl r)]^*$ unless otherwise stated.

We perform ED calculations using clusters of sizes $N=26,28,$ and 30, where the corresponding momentum clusters are depicted in Fig.~\ref{fig:nonabelian-ED-clusters}. To reduce finite-size effects, the preceding HF calculations are carried out on enlarged clusters with 234, 252, and 270 unit cells, respectively—each exactly nine times the size of the corresponding ED cluster. 

We focus on the twist angle $\theta = 2.45^\circ$, where the second moir\'e band is particularly narrow. Figures~\ref{fig:nonabelian-ED1}(a–c) presents the corresponding ED spectra. The quasi-degenerate ground states are observed in the three clusters. For the $N=28$ cluster, the ground manifold is sixfold quasi-degenerate and gapped, while the $N=26$ and $N=30$ clusters exhibit a twofold quasi-degeneracy and also gapped. This dependence of the ground-state degeneracy on the parity of the electron number $N_e' = N/2$—sixfold for even and twofold for odd—is a characteristic signature of the MR phase \cite{Read2000Paired}. Moreover, the momentum sectors in which these ground states appear agree with the predictions of the generalized Pauli principle for the MR state \cite{Haldane1991Fractional,Bernevig2012Emergent}.
Specifically, the two nearly degenerate ground states for the $N=26$ ($N=30$) cluster appear at momentum indices $0$ and $24$ ($15$), corresponding respectively to the mBZ center and the midpoint of the mBZ edge. For the $N=28$ cluster, the six quasi-degenerate ground states occur at momentum indices $7$, $14$, and $21$ (midpoints of the mBZ edges), with each sector hosting a pair of states. The correspondence between momenta and their indices is shown in Fig.~\ref{fig:nonabelian-ED-clusters}.

To further probe the nature of the states, we examine the PES, a method capable of distinguishing the MR state from competing states \cite{NonAbelian2025Liu}. The PES is constructed by dividing the system into two subsystems $A$ and $B$, containing $N_A$ and $N_B$ particles, respectively \cite{Extracting2011Sterdyniak,Chen2025Robust}. The reduced density matrix $\hat\rho_{A}$ of subsystem $A$ is defined as the  average over all reduced density matrices $\hat{\rho}_{m,A}$ for the $m$th quasi-degenerate ground state $\ket{\Psi_m}$,
\begin{align}
\hat{\rho}_A = \frac{1}{N_{\mathrm{GS}}} \sum_{m=1}^{N_{\mathrm{GS}}} \hat{\rho}_{m,A},
\end{align}
where $N_{\mathrm{GS}}$ is the quasi-degeneracy. $\hat{\rho}_{m,A}$ is obtained by tracing out the degrees of freedom of subsystem B from the density matrix $\hat\rho_m$
\begin{align}
\hat{\rho}_{m,A} = \operatorname{Tr}_B(\hat\rho_m),\hat\rho_m=\ket{\Psi_m} \bra{\Psi_m}.
\end{align}
PES is then calculated from 
\begin{align}
\hat\rho_{A}=\sum_{n}e^{-\xi_{A,n}}\ket{\alpha_{A,n}}\bra{\alpha_{A,n}},
\end{align}
where $\ket{\alpha_{A,n}}$ is the eigenstate of $\hat\rho_{A}$ and $\xi_{A,n}$ is PES in the $n$th level. We take $N_A=3$ in the calculation. Figures ~\ref{fig:nonabelian-ED1}(d-f) show PES for the quasi-degenerate states in the three clusters as a function of total momentum of particles in subsystem $A$. We find a significant entanglement gap at $\xi_{\mathrm{MR}}$ separating the spectrum in all three cases, with the low-lying levels exhibiting counting patterns that match the quasi-hole excitations for the MR state. For $N_A=3$, the counts are 2522, 3192, and 3970 for the $N=26, 28,$ and 30 clusters, respectively \cite{Read2006Wavefunctions}. Therefore, both the ED spectrum and the PES consistently indicate a MR state.

\begin{figure}[t]
   \includegraphics[width=0.5\textwidth]{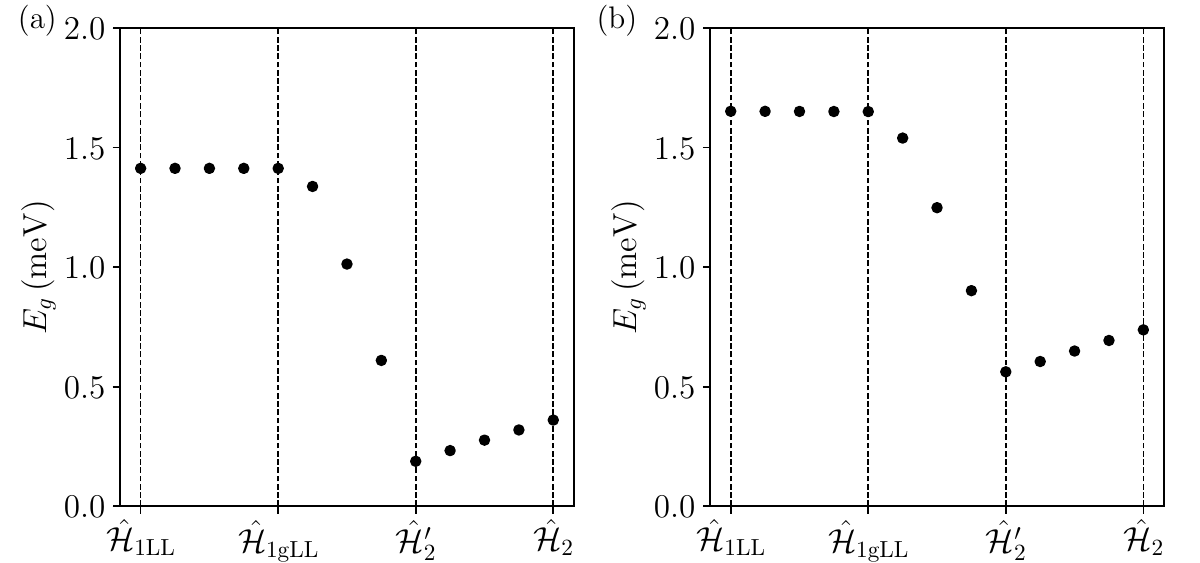}
    \caption{The energy gap $E_g$ calculated for a series of interpolated Hamiltonians between the second band of tMoTe$_2$ at $\theta=2.45^\circ$ and the first Landau level. Results are shown for (a) the $N=26$ cluster and (b) the $N=28$ cluster. The labels on the horizontal axis—$\hat{\mathcal H}_{\mathrm{1LL}}$, $\hat{\mathcal H}_{\mathrm{1gLL}}$, $\hat{\mathcal H}_{2}'$, and $\hat{\mathcal H}_2$— correspond to the Hamiltonian of the 1LL, the generalized 1LL, the tMoTe$_2$ model with zero bandwidth, and the original tMoTe$_2$ model, respectively.}
    \label{fig:adiabatic-connection}
\end{figure}

The correspondence between the second moir\'e band and a generalized first Landau level enables an adiabatic link between the MR state in tMoTe$_2$ at $\theta = 2.45^\circ$ and that in the generalized 1LL. Through this link, the state can be continuously deformed into the MR state of the standard 1LL.
To this end, we use ED to study a family of Hamiltonians that are continuously interpolated through three successive stages, with each stage controlled by a single tuning parameter.

Firstly, we set $\lambda_w=0$ and take the wave function $\wt f_{+,2,\bl{k}}(\bl{r})$ to be $[\Theta_{1,\bl k}^{(+)}(\bl r)]^*$ in Eq.~\eqref{Mkk}. We then introduce a parameter $\lambda_\mathcal{B} \in [0,1]$ to continuously interpolate the spatial fluctuations of $|\mathcal B^{(+)}(\bl r)|^2$ via:
\begin{align}
| \mathcal{B}^{(+)}(\bl r, \lambda_\mathcal{B}) |^2 = \lambda_\mathcal{B}| \mathcal{B}^{(+)}(\bl r) |^2+(1-\lambda_\mathcal{B})\mathcal{B}_0^2,
\end{align}
where $\mathcal B_0^2$ denotes the spatial average of $|\mathcal B^{(+)}(\bl r) |^2$. At $\lambda_\mathcal{B}=0$, the Hamiltonian $\hat{\mathcal H}_{\mathrm{1LL}}$ describes a flat band with Coulomb interaction projected onto the standard 1LL; at $\lambda_\mathcal{B}=1$, the Hamiltonian $\hat{\mathcal H}_{\mathrm{1gLL}}$ corresponds to a flat band with Coulomb interaction projected onto the generalized 1LL.

Secondly, we maintain $\lambda_w=0$ and $\lambda_\mathcal{B}=1$, but introduce $\lambda_F \in [0,1]$ that tunes the weight of the generalized 1LL in the wave function. The wave function $\wt f_{+,2,\bl{k}}(\bl{r})$ is taken as 
\begin{align}
\wt f_{+,2,\bl{k}}(\bl{r}) = \wt{\mathcal N}_{\bl k}\{\lambda_F[\wt{\psi}_{+,2,\bl k}(\bl r)]^*+(1-\lambda_F)[\Theta^{(+)}_{1,\bl k}(\bl r)]^*\},
\label{lambdaF}
\end{align}
where $\wt{\mathcal N}_{\bl k}$ is a normalization factor. At $\lambda_F=0$, the corresponding Hamiltonian is $\hat{\mathcal H}_{\mathrm{1gLL}}$; at $\lambda_F=1$, the Hamiltonian is denoted as $\hat{\mathcal H}_2'$. Here, we choose a gauge in which the overlap $\langle \Theta^{(+)}_{1,\bl k} | \wt{\psi}_{+,2,\bl k}\rangle$ is real and positive, ensuring that $\wt f_{+,2,\bl{k}}(\bl{r})$ varies smoothly with $\lambda_F$.  

Finally, we reintroduce the bandwidth of the second moir\'e band by tuning $\lambda_w \in [0,1]$, which allows us to interploate between $\hat{\mathcal H}_2'$ and the original Hamiltonian $\hat{\mathcal H}_2$.

Figure~\ref{fig:adiabatic-connection} presents the evolution of the many-body gap $E_g$ extracted from the ED spectra. For the $N=26$ and $N=28$ clusters, we define the gap as $E_g = E_3 - E_2$ and $E_g = E_7 - E_6$, respectively, where $E_i$ denotes the $i$th lowest energy level. The gap remains finite throughout the interpolation, establishing adiabatic continuity along the entire path. In particular, $E_g$ is nearly unchanged between $\mathcal H_{\mathrm{1LL}}$ and $\mathcal H_{\mathrm{1gLL}}$, consistent with the weak spatial modulation of the function $\mathcal B(\mathbf r)$ as shown in Fig.~\ref{fig:nonabelian2}(d). By contrast, the gap decreases rapidly as the 1LL weight is reduced from unity in $\mathcal H_{\mathrm{1gLL}}$ to $0.97$ in $\mathcal H_2'$, highlighting the strong sensitivity of the MR state to the detailed structure of the underlying wave functions. Remarkably, introducing a finite bandwidth when interpolating from $\mathcal H_2'$ to $\mathcal H_2$ slightly enhances the gap. Upon further increasing $\lambda_w$, the gap eventually closes at $\lambda_w \approx 8$ for the 26 cluster and at $\lambda_w \approx 11$ for the 30 cluster. This behavior underscores the nontrivial role of bandwidth in stabilizing non-Abelian topological states, a topic that warrants further investigation.

Given that the second moir\'e miniband resembles the 1LL, it is natural to ask whether the $\mathbb{Z}_3$ Read-Rezayi (RR) state could be stabilized. Similar to the MR state at $\nu = 5/2$, the RR state can appear at  fillings $\nu = 12/5$ and $\nu = 13/5$ in LLs and exhibit quasiparticle excitations that obey non-Abelian statistics \cite{Read1999Beyond}. Crucially, the RR state is predicted to support Fibonacci anyons, which are capable of universal topological quantum computation \cite{Nayak2008NonAbelian}. However, in our ED calculations for fillings $\nu_h = 12/5$ and $13/5$ in tMoTe$_2$ at $\theta=2.45^\circ$—corresponding to fillings $2/5$ and $3/5$ in the second moir\'e band—we do not find clear signatures of the RR phase. This is consistent with the known difficulty of stabilizing the RR state even in the standard 1LL, where it typically requires specific, finely-tuned interactions \cite{Rezayi2009NonAbelian,Wang2025Higher}. In addition, we perform ED calculation for fillings $\nu_h = 7/3$ and $8/3$ in tMoTe$_2$ at $\theta=2.45^\circ$ in a 27-unit-cell cluster, corresponding to fillings $1/3$ and $2/3$ in the second moir\'e band. The ED results, within the projected Hilbert space in Eq.~\eqref{subspace}, indicate a CDW phase at $\nu_h = 7/3$, whereas a Laughlin-type FCI is stabilized at $\nu_h = 8/3$. We note that the conventional 1LL can support Laughlin-type fractionalized states, which provide a clear example that the trace condition $\mathcal{K}=|\mathcal{C}|$ is not a necessary requirement for the existence of fractionalized states.

Motivated by the similarly large generalized 1LL weight of the second band at $\theta = 2.13^\circ$, we extend our numerical calculations to this twist angle (see Appendix~\ref{appendix:E} for details). The ED spectra and PES at $\nu_h = 5/2$ point toward a CDW state rather than a MR phase. A key factor influencing the competition between CDW and MR states appears to be the bandwidth. The phase diagram in the $(\lambda_F, \lambda_w)$ parameter space suggests that the larger bandwidth at $\theta = 2.13^\circ$ tends to destabilize the MR phase and favors the CDW state.

\section{Discussion}
\label{sec:Discussion}
In summary, we develop a theoretical framework for decomposing Chern bands into generalized Landau levels and use it to investigate Abelian and non-Abelian fractionalized states in tMoTe$_2$. Specifically, we employ a variational mapping to express the Bloch states of the first two moiré Chern bands in terms of generalized LLs, providing a clear perspective on how these fractionalized states emerge. The first moiré band is found to be dominated by the generalized 0LL, and support the formation of Abelian fractional Chern insulators in the Jain sequences, as evidenced by ED results in Figs \ref{fig:ED} and \ref{fig:ED2}.

For the second moir\'e band, HF calculations at hole filling $\nu_h=2$ reveal a dominant generalized $1$LL component at $\theta=2.13^\circ$ and $\theta=2.45^\circ$, which in turn motivates our study of interaction-driven phases at fractional fillings. Importantly, this identification is nontrivial: although the quantum weight $\mathcal{K}$ of the second band lies close to the nominal value $3|\mathcal{C}|$, such proximity alone does not a priori ensure a generalized $1$LL–like wave function. Our explicit decomposition demonstrates that resolving the internal generalized-LL structure is essential for unambiguously establishing the $1$LL-like nature of the band. ED calculations at $\nu_h=5/2$ identify the evidence of the non-Abelian MR state at $\theta=2.45^\circ$ in both the energy spectra and the PES. This MR state is further verified by its adiabatic connection to the corresponding state in the 1LL. Constructing a trial wave function for this MR state within the generalized 1LL is an interesting open problem. In contrast to the generalized 0LL, where trial wave functions for FCIs can be exactly factorized as in Eq.~\eqref{PhiF}, the generalized 1LL exhibits a more complicated structure arising from the Gram–Schmidt orthogonalization procedure used in its construction.

We also perform systematic numerical calculation at $\theta=2.13^\circ$, where a competition between the MR state and CDW state is found, with the transition primarily controlled by the bandwidth (See Appendix~\ref{appendix:E} for details). These results demonstrate a quantitative correspondence between moir\'e Chern bands and generalized LLs, establishing a foundation for understanding the emergence of fractionalized topological phases in moir\'e systems.

We compare our numerical results with available experiments. For fractionalized states at $\nu_h < 1$, our calculations predict FCIs in the Jain sequences at $\nu_h = 1/3, 2/3, 2/5, 3/5, 3/7, 4/7, 5/9$, and $4/9$ across a broad range of twist angles. Transport measurements have indeed observed fractional quantum anomalous Hall effects in tMoTe$_2$ at $\nu_h = 2/3$, $3/5$, and $4/7$ down to zero external magnetic field \cite{Park2023,Cai2023,Xu2023Observation,xu2025signatures,Xu2025Interplay,Park2025Ferromagnetism}, as well as at $\nu_h = 5/9$ under a finite magnetic field ($\sim 0.5$~T), whereas no such signatures have been reported for $\nu_h < 1/2$ \cite{xu2025signatures}. This discrepancy may arise from intrinsic effects, where fractionalized states at $\nu_h < 1/2$ are less robust or absent, or from extrinsic factors such as disorder and contact limitations at low fillings. Optical probes, however, have revealed signatures of fractionalized states even below $\nu_h = 1/2$ in the Jain sequences \cite{Pan2026Optical,Li2026Signatures}, supporting our numerical results. At $\nu_h = 5/2$, ferromagnetism has been experimentally observed \cite{Xu2025Interplay,Park2025Ferromagnetism,An2025Observation}, but evidence for non-Abelian fractionalized states remains absent. We note that disorder effects become more severe at small twist angles.  Transport and optical experiments based on high-quality devices at $\nu_h = 5/2$, combined with tuning parameters such as twist angle and pressure, could provide a decisive test of the emergence of non-Abelian fractionalized phases in tMoTe$_2$. On the numerical side, our studies employ band-projected Hamiltonians aimed at capturing the low-energy physics. Remote bands could influence the competition between fractionalized and other competing states \cite{li2025deep}, highlighting the need for further quantitative studies that go beyond the band-projection approximation. These combined theoretical and experimental efforts are essential to fully resolve the stability and nature of both Abelian and non-Abelian fractionalized phases in tMoTe$_2$ at various fillings.

More broadly, the generalized LL framework offers a systematic approach to decompose Bloch bands with arbitrary Chern numbers, as shown in Eq.~\eqref{gLLdecomp}. By examining how expansion coefficients $c_{m,n,\mathbf{k}}^{(s)}$—which control quantum geometric properties like the Chern number $\mathcal{C}$ and quantum weight $\mathcal{K}$—affect the competition between fractionalized phases and competing orders, key design principles for realizing fractionalized topological phases can be established. This could also help understand the emergence of such phases in models where bands deviate from ideal quantum geometry or are even topologically trivial \cite{Yang2025Fractional,lu2025bosonic,liu2025topological,lin2026fractional}, paving the way for a unified framework of interaction-driven fractionalized states in moir\'e and other quantum materials.

\section{Acknowledgments}
We thank Heqiu Li, Hui Liu, Yan Zhang, and Quansheng Wu for valuable discussions.
This work was supported by National Key Research and Development Program of China (Grants No. 2022YFA1402400 and No. 2021YFA1401300), National Natural Science Foundation of China (Grants No. 12274333 and No. 12550404). The numerical calculations in this paper have been performed on the supercomputing system in the Supercomputing Center of Wuhan University.

\appendix
\section{LL wave function}
\label{appendix:A} 
We present a brief review of the magnetic Bloch wave function for the $n$th LL and discuss symmetry-imposed constraints for $\mathcal{B}^{(s)}(\bl r)$. The magnetic Bloch wave function for the 0LL is given by 
 \begin{equation}
 \label{MBS0}
 \begin{aligned}
     \Psi_{0,\bl k}^{(-)}(\bl r)=&\frac{1}{S\ell}\sigma(z+iz_{\bl k}\ell^2 )e^{-\frac{1}{4}\lvert z_{\bl k}\rvert^2\ell^2-\frac{1}{4}\lvert z\rvert^2\ell^{-2}+\frac{i}{2}z_{\bl k}^*z},\\
     \Psi_{0,\bl k}^{(+)}(\bl r)=&[\Psi_{0,-\bl k}^{(-)}(\bl r)]^*,
 \end{aligned}     
 \end{equation}
where $z = x + iy, z_{\bl{k}} = k_x + ik_y$ , $S$ is a normalization factor, $\ell = \sqrt{\mathcal{A}_0/(2\pi)}$,  $\mathcal{A}_0$ is the area of the (magnetic) unit cell, and $\sigma(z)$ is the modified Weierstrass sigma function \cite{Haldane2018modular} formulated as
\begin{equation}
\begin{aligned}
\label{sigma}
\sigma(z)= ze^{\frac{\eta_1z^2}{z_1}}\frac{\mathcal{\theta}_1(v\mid\tau)}{v\mathcal{\theta}_1'(0\mid\tau)},
\end{aligned}
\end{equation}
where $\mathcal{\theta}_1(v\mid\tau)$ is the Jacobi theta function, $v=z/z_1$, $\eta_1=z_1^*/(4\ell^2)$, $\tau=z_2/z_1$, and $z_j=a_{j,x}+ia_{j,y}$. Here $\bl a_{1,2}$ are  primitive (magnetic) lattice vectors, and we take $\bl a_{1,2}=(\pm\frac{\sqrt{3}}{2},\frac{1}{2})a_M$. Expression of $\mathcal{\theta}_1(u\mid\tau)$ is
\begin{align}
\mathcal{\theta}_{1}(u\mid\tau)=-\sum_{n=-\infty}^{+\infty}e^{i\pi\tau(n+\frac{1}{2})^2}e^{2\pi i(n+1/2)(u+1/2)}.
\end{align}
The magnetic Bloch wavefunction for the $n$th LL is formulated as
\begin{equation}
 \label{MBSn}
 \begin{aligned}
     \Psi_{n,\bl k}^{(-)}(\bl r)=&\frac{(a^\dagger)^n}{\sqrt{n!}}\Psi_{0,\bl k}^{(-)}(\bl r),\\
     \Psi_{n,\bl k}^{(+)}(\bl r)=&[\Psi_{n,-\bl k}^{(-)}(\bl r)]^*,
 \end{aligned}     
 \end{equation}
where $a^\dagger$ and $a$ denote, respectively, the raising and lowering operators  
\begin{equation}
\begin{aligned}
&a^\dagger=i\frac{-2\ell\partial_z+z^*\ell^{-1}/2}{\sqrt2},a=i\frac{-2\ell\partial_{z^*}-\ell^{-1}z/2}{\sqrt2}. 
\end{aligned}
\end{equation} 
$\Psi_{n,\boldsymbol k}^{(s)}(\boldsymbol r)$ satisfies the following magnetic translational symmetry and point-group symmetry \cite{li2025Variational}
\begin{equation}
 \label{sym_Psi}
 \begin{aligned}
&\Psi_{n,\boldsymbol k}^{(s)}(\boldsymbol r+\boldsymbol{a}_i)
=-e^{-is\frac{1}{2\ell^2}\boldsymbol {a}_i\times \boldsymbol r}e^{i\boldsymbol k\cdot \boldsymbol {a}_i}\Psi^{(s)}_{n,\boldsymbol k}(\boldsymbol r),\\
&\Psi^{(s)}_{0,\boldsymbol \gamma}(\hat R_{3z}\boldsymbol r)=e^{-i\frac{2}{3}\pi s}\Psi^{(s)}_{0,\boldsymbol \gamma}(\boldsymbol r),\\
&\Psi^{(s)}_{0,\boldsymbol{\kappa}_\pm}(\hat R_{3z}\boldsymbol r)=\Psi^{(s)}_{0,\boldsymbol{\kappa}_\pm}(\boldsymbol r),\\
&\Psi^{(s)}_{0,\boldsymbol\gamma}(\hat R_{2y}\boldsymbol r)=-[\Psi^{(s)}_{0,\boldsymbol\gamma}(\boldsymbol r)]^*,
\end{aligned}     
\end{equation}
where $\hat{R}_{ni}$ represents the $n$-fold rotation around the $i$-axis.

We now derive the constraints on $\mathcal B^{(s)}(\bl r)$ imposed by the symmetry properties of $\gLLsr{s}{n}$. $\gLLsr{s}{n}$ is the Bloch wave function satisfying the translational symmetry
\begin{equation}
\label{sym_trans_theta}
\begin{aligned}
&\Theta_{n,\bl k}^{(s)}(\bl r+\bl a_i)=e^{i\bl k\cdot\bl a_i}\gLLsr{s}{n},
 \end{aligned}     
 \end{equation} 
and has a decomposition \begin{align}
\gLLsr{s}{n}=\mathcal B^{(s)}(\bl r)\xi_{\bl k}(\bl r),
\end{align}
where $\xi_{\bl k}(\bl r)$ is a linear combination of $\Psi_{m,\boldsymbol k}^{(s)}(\boldsymbol r)$. As a consequence, $\mathcal B^{(s)}(\bl r)$ is constrained by the translational symmetry,
\begin{align}
&\mathcal B^{(s)}(\bl r+\bl{a}_i) = -e^{is\frac{1}{2\ell^2}\boldsymbol {a}_i\times \boldsymbol r}\mathcal B^{(s)}(\bl r).
\end{align}

In the case of tMoTe$_2$, The $C_{3z}$ and $C_{2y}\mathcal{T}$ symmetries of the moir\'e Hamiltonian $H_+$ can be represented by
\begin{equation}
\begin{aligned}
\hat C_{3z} = & U_0(\bl r)\hat R_{3z}U_0^\dagger(\bl r),\hat C_{2y}\hat{\mathcal{T}} = \sigma_x\hat R_{2y}\hat{\mathcal{T}}.
\end{aligned}     
\end{equation}
Under $C_{3z}$ and $C_{2y}\mathcal{T}$ symmetries, the Hamiltonian of tMoTe$_2$ transforms as follows,
\begin{equation}
\begin{aligned}
&\hat C_{3z}H_+(\bl k,\bl r)\hat C_{3z}^{-1}=H_+(\bl k,\bl r),\\
&[\hat C_{2y}\hat{\mathcal{T}}]H_+(\bl k,\bl r)[\hat C_{2y}\hat{\mathcal{T}}]^{-1}=H_+(\bl k,\bl r).
\end{aligned}     
\end{equation}

Since $\gLLsr{s}{n}$ are basis functions to decompose Bloch states in tMoTe$_2$, we require that they are eigenstates of $\hat C_{3z}$ at the threefold rotation invariant momentum $\{\bl \kappa_\pm,\bl \gamma\}$, and also eigenstate of $\hat C_{2y}\hat{\mathcal{T}}$ at $\bl \gamma$ point. Constrained by these symmetries, $\gLLsr{s}{n}$ satisfies the following transformation rules:
\begin{equation}
 \label{sym_theta}
 \begin{aligned}
 &\hat C_{3z}\Theta_{n,\bl k}^{(s)}(\bl r)=e^{i\frac{2\pi}{3}L_{s,n,\bl k}}\Theta_{n,\bl k}^{(s)}(\bl r),\\
&\hat C_{2y}\hat{\mathcal{T}}\Theta_{n,\bl \gamma}^{(s)}(\bl r)=e^{i\pi M_{s,n,\bl \gamma}}\Theta_{n,\bl \gamma}^{(s)}(\bl r),
 \end{aligned}     
 \end{equation}
 where $\bl k\in\{\bl \kappa_\pm,\bl \gamma\}$ are three high-symmetry points in mBZ for the first line. Here $L_{s,n,\bl k} \in \{-1,0,1\}$ and $M_{s,n,\bl \gamma} \in \{0,1\}$ are integers that label the eigenvalues of $C_{3z}$ and $C_{2y}\mathcal{T}$ symmetries, respectively. Therefore, $\mathcal B^{(s)}(\bl r)$ must satisfy,
\begin{align}
\label{Br_sym1}
&\hat C_{3z}\mathcal B^{(s)}(\bl r)=e^{iv_s}\mathcal B^{(s)}(\bl r),\\
\label{Br_sym2}
&\hat C_{2y}\hat{\mathcal{T}}\mathcal B^{(s)}(\bl r)=e^{iw_s}\mathcal B^{(s)}(\bl r),
\end{align}
where $v_+$ and $w_+$ are determined by the ansatz for $\mathcal B^{(+)}(\bl r)$ given in Eq.~\eqref{Br0}. 

We now demonstrate that $\mathcal{B}^{(-)}(\bm r)$ in Eq.~\eqref{Br_relation} satisfies Eqs.~\eqref{Br_sym1} and \eqref{Br_sym2}. For Eq.~\eqref{Br_sym1}, we have
\begin{equation}
\begin{aligned}
&\hat C_{3z}\mathcal{B}^{(-)}(\bl r)\\
=&U_0(\bl r)i\sigma_y[U_0^\dagger(\hat R_{3z}\bl r)\mathcal B^{(+)}(\hat R_{3z}\bl r)]^*\\
=&U_0(\bl r)i\sigma_y[U_0^\dagger(\bl r)\hat C_{3z}\mathcal B^{(+)}(\bl r)]^*\\
=&e^{-iv_+}U_0(\bl r)i\sigma_y[U_0^\dagger(\bl r)\mathcal B^{(+)}(\bl r)]^*\\
=& e^{-iv_+}\mathcal B^{(-)}(\bl r).
\end{aligned}   
\end{equation}
Hence, Eq.~\eqref{Br_sym1} is satisfied and $v_- = -v_+$. For Eq.~\eqref{Br_sym2}, we have 
\begin{equation}
\begin{aligned}
&\hat C_{2y}\hat{\mathcal{T}}\mathcal B^{(-)}(\bl r)\\
=&-e^{-i(\bl\kappa_++\bl\kappa_-)\cdot\hat R_{2y}\bl r}i\sigma_y\sigma_x[\mathcal{B}^{(+)}(\hat R_{2y}\bl r)]^*
 \\
=&e^{-iw_++i\pi}\mathcal B^{(-)}(\bl r).
\end{aligned}   
\end{equation}
Therefore Eq.~\eqref{Br_sym2} is satisfied and we obtain $w_- = -w_+ + \pi$.

\section{QGT of generalized LL}
\label{appendix:B}
We now derive the QGT of the generalized LL wave function and prove the integrated form of the trace condition in Eq.~\eqref{trace-condition2}. For simplicity, we focus on the orientation of magnetic field $s=-$, and the expressions for the $s=+$ case can be derived similarly. The QGT of the generalized LL wave function is determined by the Berry connection from equation
\begin{equation}
\label{QGT_gLL_from_Berry}
\begin{aligned}
& (\chi_{n,\bl k}^{(\mathrm{gLL})})_{ij}  \\
= & \bra{\partial_{k_i}\gLLu{n}}{\partial_{k_j}\gLLu{n}}\rangle- \bra{\partial_{k_i}\gLLu{n}}\gLLu{n}\rangle\langle\gLLu{n}|{\partial_{k_j}\gLLu{n}}\rangle \\
= & \sum_{m\ne n}A_{m,n,i}^*A_{m,n,j},
\end{aligned}
\end{equation}
where $i, j$ are coordinate labels, and $\gLLur{n}=e^{-i\bl k\cdot\bl r}\Theta_{n,\bl k}^{(-)}(\bl r)$. The Berry connection $A_{m,n,j}$ is given by 
\begin{equation}
\label{Berry_connection}
\begin{aligned}
A_{m,n,j}= i\braket{\gLLu{m}}{\partial_{k_j}\gLLu{n}}.
\end{aligned}
\end{equation}

The partial derivative operator $\partial_{k_j}$ can be rewritten using the raising and lowering operators, $b^\dagger$  and $b$,  in momentum space
\begin{equation}
\begin{aligned}
\partial_{k_x}= & \partial_{z_{\bl k}}+\partial_{z_{\bl k}^*}=\ell\frac{b-b^\dagger}{\sqrt{2}}-\frac{i\ell^2k_y}{2},\\
\partial_{k_y}= & i(\partial_{z_{\bl k}}-\partial_{z_{\bl k}^*})=-i\ell\frac{b+b^\dagger}{\sqrt{2}}+\frac{i\ell^2k_x}{2}.
\end{aligned}
\end{equation}
Here $b$ and $b^\dagger$ are obtained from $a$ and $a^\dagger$ by substituting $z \rightarrow i\ell^2 z_{\boldsymbol{k}}$,
\begin{equation}\begin{aligned}
&b^\dagger=\frac{-2\ell^{-1}\partial_{z_{\bl k}}+\ell z_{\bl k}^*/2}{\sqrt2},b=\frac{2\ell^{-1}\partial_{z_{\bl k}^*}+\ell z_{\bl k}/2}{\sqrt2}.
\end{aligned}\end{equation}

For the conventional LL wavefunction, we have the recurrence relation in the momentu space, 
\begin{equation}
 \label{MBSn-v2}
 \begin{aligned}
\wt{\Psi}_{n,\bl k}^{(-)}(\bl r)=\frac{(b^\dagger)^n}{\sqrt{n!}}\wt{\Psi}_{0,\bl k}^{(-)}(\bl r),
 \end{aligned}     
 \end{equation}
where $\wt{\Psi}_{n,\bl k}^{(-)}(\bl r)=e^{-i\bl k\cdot\bl r}\Psi_{n,\bl k}^{(-)}(\bl r)$.

We define the matrix element of  $b^\dagger$ and $b$ operators in the generalized LL basis,
\begin{equation}
\begin{aligned}
\gamma_{m,n,\bl k}= \braket{\gLLu{m}}{b^\dagger\gLLu{n}},\gamma_{m,n,\bl k}^\prime= \braket{\gLLu{m}}{b\gLLu{n}},
\end{aligned}
\end{equation}
which are related by $\gamma_{m,n, \bl k}^\prime=\gamma_{n,m,\bl k}^*$. 
We note that the inner products used throughout this work are defined in real space, while the momentum index acts as a quantum number.
The Berry connection can be expressed using these matrix elements as,
\begin{equation}
\begin{aligned}
A_{m,n,x}= & i\frac{\ell}{\sqrt{2}}(\gamma_{n,m,\bl k}^*-\gamma_{m,n,\bl k})+\frac{\ell^2}{2}k_y\delta_{m,n}\\
A_{m,n,y}= & \frac{\ell}{\sqrt{2}}(\gamma_{n,m,\bl k}^*+\gamma_{m,n,\bl k})-\frac{\ell^2}{2}k_x\delta_{m,n}.
\end{aligned}
\end{equation}

We begin by proving that  $\gamma_{m,n,\bl k}$ is nonzero only when $m=n$ and $m=n+1$. For a linear combination of $\gLLur{m}$ denoted by $f(\bl r)$, we have decomposition
\begin{equation}
\begin{aligned}
f(\bl r)= &\mathcal B^{(-)}(\bl r)g(\bl r),
\end{aligned}
\end{equation}
where $g(\bl r)$ is a linear combination of $\wt{\Psi}_{a,\bl k}^{(-)}(\bl r)$ with a maximal LL index $a_{\text{max}}$. Specifically, $f(\bl r)=\gLLur{n}$ and $f(\bl r)=b^\dagger\gLLur{n}$ correspond to $a_{\text{max}}=n$ and $n+1$, respectively. Therefore, $b^\dagger\gLLur{n}$ can be decomposed as,
\begin{equation}
\begin{aligned}
\label{decomp}
&b^\dagger \gLLur{n}=\sum_{m=0}^{n+1}\gamma_{m,n, \bl k}\gLLur{m},
\end{aligned}
\end{equation}
which indicates that $\gamma_{m,n,\bl k}=0$ for $m>n+1$. Similarly, we have 
\begin{equation}
\begin{aligned}
\label{decomp2}
b\gLLur{n} = \sum_{m=0}^{n}\gamma_{m,n,\bl k}^\prime\gLLur{m}.
\end{aligned}
\end{equation}
Therefore, $\gamma_{m,n, \bl k}=[\gamma_{n,m,\bl k}^\prime]^*=0$ for $m<n$.

We first determine $\gamma_{n+1,n, \bl k}$ through the following equation,
\begin{equation}
\begin{small}
\begin{aligned}
\label{n_n-1_is_real}
& \gamma_{n+1,n,\bl k} =  \braket{\gLLu{n+1}}{b^\dagger \mathcal N_{n,\bl k} \wt e_{n,\bl k}}  \\
= & \sqrt{n+1}\mathcal N_{n,\bl k}\braket{\gLLu{n+1}}{\wt e_{n+1,\bl k}} =  \sqrt{n+1} \frac{\mathcal N_{n,\bl k}}{\mathcal N_{n+1,\bl k}},
\end{aligned}
\end{small}
\end{equation}
where $\wt e_{n,\bl k}(\bl r)=e^{-i\bl k\cdot\bl r}e_{n,\bl k}(\bl r)$. We then obtain a recurrence relation for $\gamma_{n,n, \bl k}$ through the following relations,
\begin{equation}
\begin{small}
\begin{aligned}
&\braket{b^\dagger \wt{\Theta}_{n,\bl k}^{(-)}}{b^\dagger\wt{\Theta}_{n+1,\bl k}^{(-)}}\\
=&-\sqrt{2}\ell^{-1}\partial_{z_{\bl k}}\gamma_{n+1,n, \bl k}+\braket{b \wt{\Theta}_{n,\bl k}^{(-)}}{b\wt{\Theta}_{n+1,\bl k}^{(-)}},
\end{aligned}
\end{small}
\end{equation} 
\begin{equation}
\begin{aligned}
&\braket{b^\dagger \wt{\Theta}_{n,\bl k}^{(-)}}{b^\dagger\wt{\Theta}_{n+1,\bl k}^{(-)}} \\=&
\braket{b^\dagger \wt{\Theta}_{n,\bl k}^{(-)}}{\wt{\Theta}_{n+1,\bl k}^{(-)}}
\braket{\wt{\Theta}_{n+1,\bl k}^{(-)}}{b^\dagger\wt{\Theta}_{n+1,\bl k}^{(-)}}\\
=& \gamma_{n+1,n,\bl k}\gamma_{n+1,n+1, \bl k},
\end{aligned}
\end{equation}
\begin{equation}
\begin{aligned}
&\braket{b \wt{\Theta}_{n,\bl k}^{(-)}}{b\wt{\Theta}_{n+1,\bl k}^{(-)}} \\=&
\braket{b \wt{\Theta}_{n,\bl k}^{(-)}}{\wt{\Theta}_{n,\bl k}^{(-)}}
\braket{\wt{\Theta}_{n,\bl k}^{(-)}}{b\wt{\Theta}_{n+1,\bl k}^{(-)}}\\
=& \gamma_{n+1,n, \bl k}\gamma_{n,n,\bl k}.
\end{aligned}
\end{equation}
The recurrence relation is
\begin{equation}
\begin{small}
\begin{aligned}
\label{recur_nn}
&\gamma_{n+1,n+1, \bl k}= -\sqrt{2}\ell^{-1}\partial_{z_{\bl k}}\ln\gamma_{n+1,n, \bl k}+\gamma_{n,n, \bl k}.
\end{aligned}
\end{small}
\end{equation}

Combining Eq.~\eqref{n_n-1_is_real} with the recurrence relation in Eq.~\eqref{recur_nn}, we obtain the expression for $\gamma_{n,n,\bl k}$
\begin{equation}
\begin{small}
\begin{aligned}
\label{eq_nn}
&\gamma_{n,n,\bl k}=\sqrt{2}\ell^{-1}\partial_{z_{\bl k}}\ln\mathcal{N}_{n,\bl k}.
\end{aligned}
\end{small}
\end{equation}

The Berry connection has the following form,
\begin{equation} 
\label{Ax}
\begin{small} 
\begin{aligned}
A_{m,n,x}
= &\begin{cases}
-\frac{i}{\sqrt{2}}\ell\gamma_{n+1,n,\bl k}& m=n+1\\
-\partial_{k_y}\ln\mathcal N_{n,\bl k}+\frac{\ell^2k_y}{2}&m=n\\
\frac{i}{\sqrt{2}}\ell\gamma_{n,n-1,\bl k}& m=n-1\\
0&m\ne n+1,n,n-1
\end{cases},
\end{aligned}
\end{small} 
\end{equation}
and
\begin{equation} 
\label{Ay}
\begin{small} 
\begin{aligned}
A_{m,n,y}
= &\begin{cases}
\frac{1}{\sqrt{2}}\ell\gamma_{n+1,n,\bl k}& m=n+1\\
\partial_{k_x}\ln\mathcal N_{n,\bl k}-\frac{\ell^2k_x}{2}&m=n\\
\frac{1}{\sqrt{2}}\ell\gamma_{n,n-1,\bl k}& m=n-1\\
0&m\ne n+1,n,n-1
\end{cases}.
\end{aligned} 
\end{small} 
\end{equation}

After substituting Eqs.~\eqref{Ax} and \eqref{Ay} into Eq.~\eqref{QGT_gLL_from_Berry}, we derive the QGT for the generalized LL wave functions. 

For $n=0$, $\chi^{(\mathrm{gLL})}_{0,\bl k}$ is given by
\begin{equation}
\begin{small}
\begin{aligned}
\chi^{(\mathrm{gLL})}_{0,\bl k}=
\gamma_{1,0,\bl k}^2\ell^2\begin{pmatrix} 
 \frac{1}{2} &  \frac{i}{2}
 \\ 
-\frac{i}{2} &  \frac{1}{2}
\end{pmatrix},
\end{aligned}
\end{small}
\end{equation}
which clearly satisfies the trace condition $\mathrm{Tr}[g_{\bl k}]=\abs{\Omega_{\bl k}}$.

For $n\ge 1$, $\chi^{(\mathrm{gLL})}_{n,\bl k}$ is formulated as 
\begin{equation}
\begin{small}
\begin{aligned}
\label{QGT}
\chi^{(\mathrm{gLL})}_{n,\bl k}=\ell^2\begin{pmatrix} 
 \frac{1}{2}[\gamma_{n+1,n,\bl k}^2+\gamma_{n,n-1,\bl k}^2]&  
\frac{i}{2}[\gamma_{n+1,n,\bl k}^2-\gamma_{n,n-1,\bl k}^2]
 \\ 
-\frac{i}{2}[\gamma_{n+1,n,\bl k}^2-\gamma_{n,n-1,\bl k}^2]
 &  
 \frac{1}{2}[\gamma_{n+1,n,\bl k}^2+\gamma_{n,n-1,\bl k}^2]
\end{pmatrix}.
\end{aligned}
\end{small}
\end{equation} 

We note that the Berry curvature can also be formulated as,
\begin{equation}
\begin{aligned}
\Omega_{n,\bl k}=-2\mathrm{Im}[\chi^{(\mathrm{gLL})}_{n,\bl k}]=\partial_{k_x}A_{n,n,y}-\partial_{k_y}A_{n,n,x},
\end{aligned}
\end{equation}
from which we derive the recurrence relation for $\gamma_{n+1,n,\bl k}$
\begin{equation}
\begin{small}
\begin{aligned}
\label{recur_n_n-1}
\gamma_{n+1,n,\bl k}^2-\gamma_{n,n-1,\bl k}^2
=-4\ell^{-2}\partial_{z_{\bl k}}\partial_{z_{\bl k}^*}\ln\mathcal N_{n,\bl k}+1.
\end{aligned}
\end{small}
\end{equation}
Combining Eqs.~\eqref{n_n-1_is_real} and \eqref{recur_n_n-1}, we obtain the recurrence relation for $\mathcal N_{n,\bl k}$
\begin{equation}
\begin{small}
\begin{aligned}
(n+1)\frac{\mathcal N_{n,\bl k}^2}{\mathcal N_{n+1,\bl k}^2}-n\frac{\mathcal N_{n-1,\bl k}^2}{\mathcal N_{n,\bl k}^2}
=-4\ell^{-2}\partial_{z_{\bl k}}\partial_{z_{\bl k}^*}\ln\mathcal N_{n,\bl k}+1.
\end{aligned}
\end{small}
\end{equation}
This equation also applies to $n=0$ with the understanding that $\mathcal N_{-1,\bl k}=0$. Since $\mathcal N_{n,\bl k}$ is a continuous periodic function of $\bl k$, we establish the following formula 
\begin{equation}
\begin{aligned}
\label{int_condition}
\frac{1}{2\pi}\int d^2\bl k\,\gamma_{n+1,n,\bl k}^2=\ell^{-2}+\frac{1}{2\pi}\int d^2\bl k\,\gamma_{n,n-1,\bl k}^2.
\end{aligned}
\end{equation}
Combining Eqs.~\eqref{QGT} and \eqref{int_condition}, we obtain the integrated form of the trace condition in Eq.~\eqref{trace-condition2}.

\begin{figure}[t]   \includegraphics[width=1.\columnwidth]{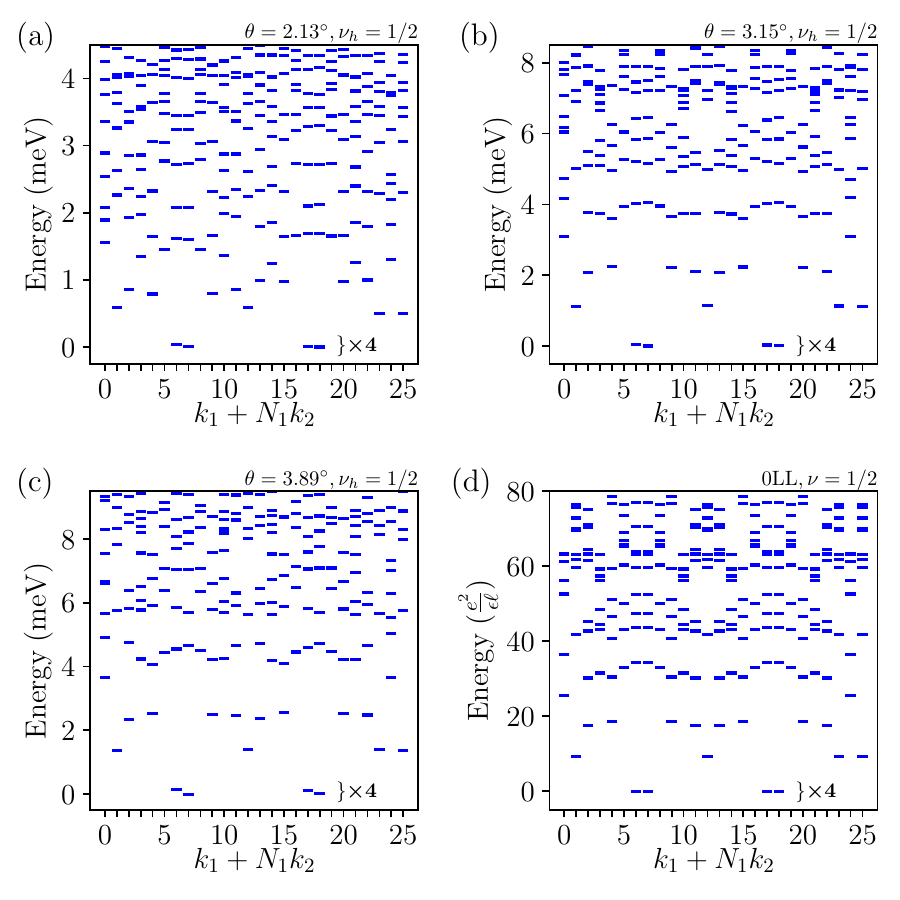}
    \caption{(a-c) ED spectra of the original tMoTe$_2$ model at $\nu_h=1/2$, where the twist angles are $\theta=2.13^\circ,3.15^\circ$, and $3.89^\circ$. (d) ED spectrum of the 0LL at $\nu=1/2$. The energy scale is $e^2/(\epsilon \ell)$ with $\ell$ being the magnetic length. A 26 momentum cluster is employed.} 
    \label{fig:ED-appendix}
\end{figure}

\begin{figure*}[t]   \includegraphics[width=\textwidth]{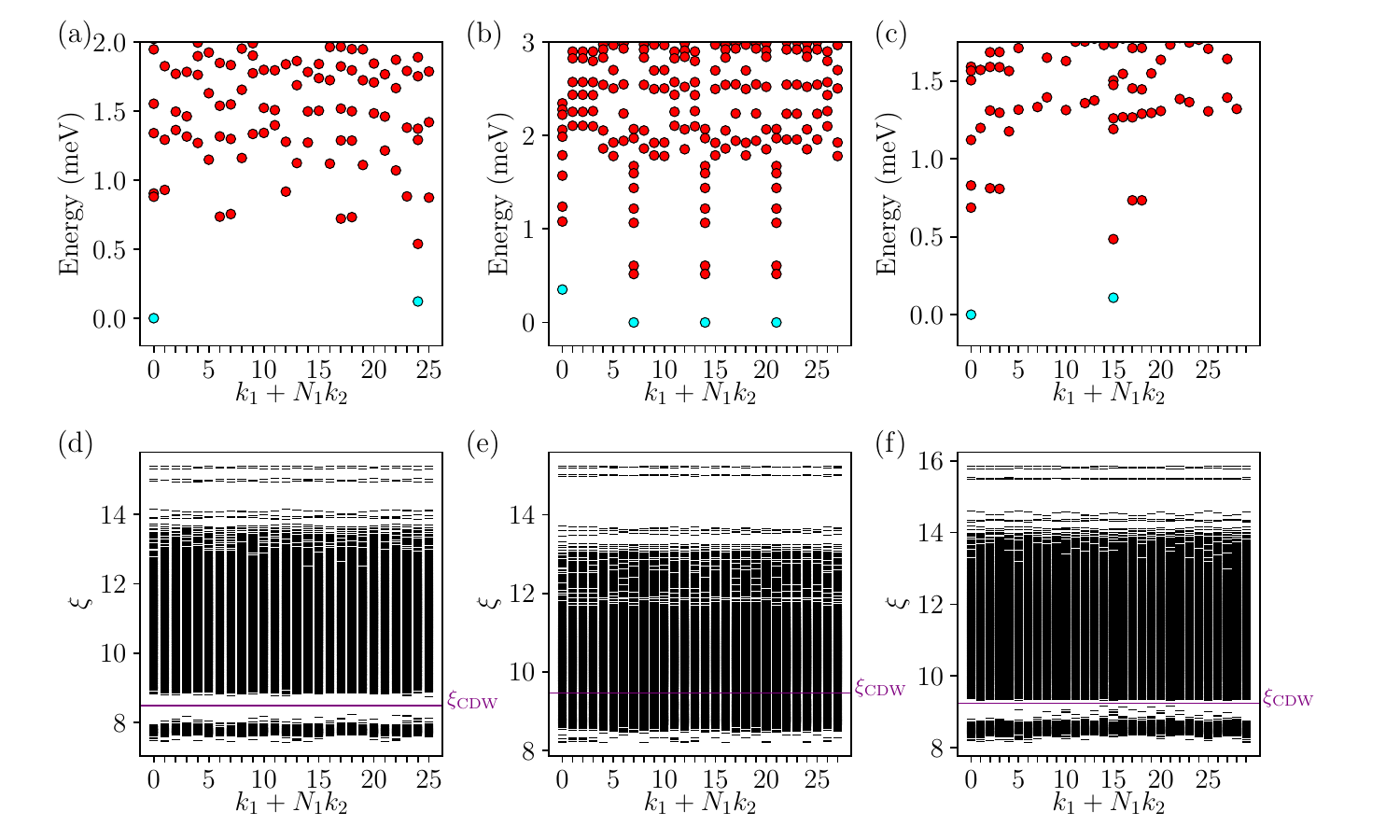}
    \caption{(a-c) ED spectrum obtained at $\nu_h=5/2$ and $\theta = 2.13^\circ$, using $N=26,28$, and 30 clusters. The quasi-degenerate states are highlighted in cyan. (d-f) PES with $N_A=4$, for the quasi-degenerate states highlighted in cyan in (a-c).}
    \label{fig:nonabelian-ED-2.13}
\end{figure*}

\section{self-consistent HF method}
\label{appendix:C}
We present a detailed description of the self-consistent HF method. The full many-body Hamiltonian of in tMoTe$_2$ in the hole basis is given by
\begin{equation}
\begin{aligned}
\label{full}
&\hat H^{\mathrm{(full)}}=  \hat H_1+\hat H_2\\
\end{aligned}
\end{equation}
The single-particle Hamiltonian $\hat H_1$ is 
\begin{equation}
\begin{aligned}
&\hat{H}_{1} = \sum_{\tau,n,\bl k}(-\mathcal{E}_{\tau,n,\bl k})\varphi^{\dagger}_{\tau,n,\bl k}\varphi_{\tau,n,\bl k},\\
\end{aligned}
\end{equation}
where $-\mathcal{E}_{\tau,n,\bl k}$ is the single-particle energy in the hole basis. The interaction term $\hat H_2$ is 
\begin{equation}
\begin{aligned}
&\hat{H}_2 =  \sum_{\tau,\tau'}\sum_{n_1,n_2,n_3,n_4}\sum_{\bl k_1,\bl k_2,\bl k_3,\bl k_4} V_{\bl k_1\bl k_2\bl k_3\bl k_4}^{\tau\tau'n_1n_2n_3n_4} \\
&\;\;\;\;\;\;\times \varphi^{\dagger}_{\tau,n_1,\bl k_1}\varphi^{\dagger}_{\tau',n_2,\bl k_2} \varphi_{\tau',n_3,\bl k_3}\varphi_{\tau,n_4,\bl k_4}.
\end{aligned}
\end{equation}
The interaction matrix element $V^{\tau\tau'n_1n_2n_3n_4}_{\bl{k_1k_2k_3k_4}}$ is given by
\begin{equation}
\begin{aligned}
 V^{\tau\tau'n_1n_2n_3n_4}_{\bl{k_1k_2k_3k_4}}=&\frac{1}{2\mathcal{A}}\sum_{\bl{q}}V(\bl q)
    M^{\tau n_1n_4}_{\bl{k_1}\bl{k_4}}(\bl{q})
    M^{\tau' n_2n_3}_{\bl{k_2}\bl{k_3}}(\bl{-q}),
\end{aligned}
\label{IntV}
\end{equation}
where $M$ is the plane-wave matrix element.

We then present the self-consistent HF method in the plane-wave basis.
The hole creation (annihilation) operator in the plane-wave basis, $b_{\tau,\bl k+\bl g,l}^{\dagger}$ ($b_{\tau,\bl k+\bl g,l}$), are defined from
\begin{equation}
\begin{aligned}
&\varphi^\dagger_{\tau,n,\bl k}=\sum_{\bl g,l}U_{\tau,n,\bl k+\bl g,l}^*b^{\dagger}_{\tau,\bl k+\bl g,l}\\
&\varphi_{\tau,n,\bl k}=\sum_{\bl g,l}U_{\tau,n,\bl k+\bl g,l}b_{\tau,\bl k+\bl g,l}
\end{aligned}   
\end{equation} 
where $\bl g$ represents the basis vector in mBZ and $U_{\tau,n,\bl k+\bl g,l}$ satisfies 
\begin{equation}
\begin{aligned}
[h_{\tau,\bl k}]_{\bl gl,\bl g'l'}=\sum_nU_{\tau,n,\bl k+\bl g,l}\mathcal{E}_{\tau,n,\bl k}U_{\tau,n,\bl k+\bl g',l}^*.
\end{aligned}   
\end{equation} where $h_\tau$ is the Hamiltonian of tMoTe$_2$ in the plane-wave basis. $\hat H_1$ and $\hat H_2$ are rewritten as
\begin{equation}
\begin{aligned}
\label{full_plane_wave}
&\hat{H}_{1} = -\sum_{\bl k,\bl g,\bl g'}\sum_{l,l'}\sum_{\tau}[h_{\tau,\bl k}^{\mathsf{T}}]_{\bl gl,\bl g'l'}b^{\dagger}_{\tau,\bl k+\bl g,l}b_{\tau,\bl k+\bl g',l'}\\
&\hat{H}_2 = \frac{1}{2\mathcal A} \sum_{\bl k,\bl k',\bl q}\sum_{\bl g,\bl g'} \sum_{l,l',\tau,\tau'} V(\bl q) \\
&\;\;\;\;\;\;\times b^{\dagger}_{\tau,\bl k+\bl g+\bl q,l} b^{\dagger}_{\tau',\bl k'+\bl g'-\bl q,l'} b_{\tau',\bl k'+\bl g',l'} b_{\tau,\bl k+\bl g,l}.\\
\end{aligned}
\end{equation}

Applying the HF approximation to $\hat H^{\mathrm{(full)}}$, we obtain the mean-field Hamiltonian,
\begin{equation}
\begin{small}
\begin{aligned}
&\hat{H}^{(\mathrm{HF})} = \\
&\sum_{\bl k,\bl g,\bl g'}\sum_{l,l'}\sum_{\tau}[-h_{\tau,\bl k}^{\mathsf{T}}+V^{(\mathrm{HF})}_{\tau,\bl k}]_{\bl gl,\bl g'l'}b^{\dagger}_{\tau,\bl k+\bl g,l}b_{\tau,\bl k+\bl g',l'}+E_0
\end{aligned}
\end{small}
\end{equation}
where $V^{(\mathrm{HF})}_{\tau,\bl k}$ is the HF correction term given by 
\begin{equation}
\begin{aligned}
&[V^{(\mathrm{HF})}_{\tau,\bl k}]_{\bl gl,\bl g'l'} \\
= & \delta_{ll'}\frac{1}{\mathcal A}  \sum_{\bl k',\bl g''}\sum_{l'',\tau'}V(\bl g-\bl g')[n_{\tau',\bl k'}]_{(\bl g'+\bl g'')l'',(\bl g+\bl g'')l''} \\
-& \frac{1}{\mathcal A} \sum_{\bl k',\bl g''} V(\bl k'-\bl k+\bl g'')[n_{\tau,\bl k'}]_{(\bl g'+\bl g'')l',(\bl g+\bl g'')l}.
\end{aligned}
\end{equation}
and the constant energy term $E_0$ is 
\begin{align}
    E_0=&-\frac{1}{2}\sum_{\bl k,\bl g,\bl g'}\sum_{l,l'}\sum_{\tau}[V^{(\mathrm{HF})}_{\tau,\bl k}]_{\bl gl,\bl g'l'}[n_{\tau,\bl k}]_{\bl gl,\bl g'l'}.
\end{align}
$n_{\tau,\bl k}$ is the density matrix
\begin{align}
[n_{\tau,\bl k}]_{\bl gl,\bl g'l'}=\bra{\Omega}b^{\dagger}_{\bl k+\bl g,l,\tau} b_{\bl k+\bl g',l',\tau}\ket{\Omega}.
\end{align}
Here $\ket{\Omega}$ denotes the ground state with the first band fully occupied when the filling factor $\nu_h=2$ is considered,
\begin{align}
\ket{\Omega}=&\prod_{\tau=\pm}\prod_{i=1}^N\varphi^\dagger_{\tau,1,\bl k_i}\ket{0}\\
=&\prod_{\tau=\pm}\prod_{i=1}^N\sum_{\bl g,l}U_{\tau,1,\bl k_i+\bl g,l}^*b^{\dagger}_{\tau,\bl k_i+\bl g,l}\ket{0},
\end{align} 
Using a plane-wave cutoff of $\lvert\mathbf{g}\rvert \le 5\lvert\mathbf{g}_1\rvert$, we carry out the self-consistent Hartree-Fock calculation iteratively. In each step, we diagonalize $\hat{H}^{(\mathrm{HF})}$ and update $\ket{\Omega}$ with the resulting eigenstates, repeating until convergence is reached. This minimizes the total energy $E^{(\mathrm{HF})}$ of the filled first band, which is expressed as
\begin{align}
&E^{(\mathrm{HF})}=\bra{\Omega}\hat{H}^{(\mathrm{HF})}\ket{\Omega}\\
&=\sum_{\bl k,\bl g,\bl g'}\sum_{l,l'}\sum_{\tau}[-h_{\tau,\bl k}^{\mathsf{T}}+\frac{1}{2}V^{(\mathrm{HF})}_{\tau,\bl k}]_{\bl gl,\bl g'l'}[n_{\tau,\bl k}]_{\bl gl,\bl g'l'},
\end{align}
and the convergence is reached when the change in the Hartree–Fock energy between successive iterations falls below $10^{-5}$ meV.

\begin{figure}[t]
   \includegraphics[width=0.5\textwidth]{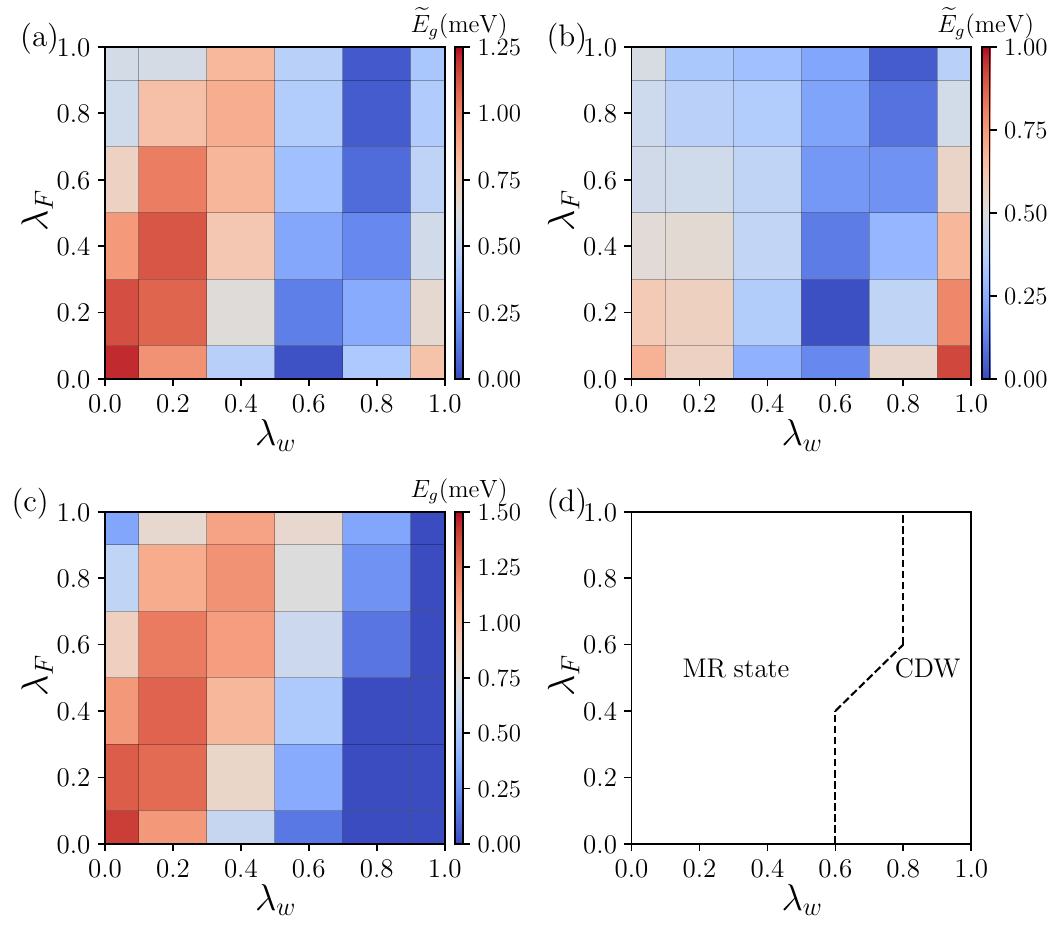}
    \caption{Maps of the energy gap for quasi-degenerate states in the ($\lambda_F,\lambda_w$) parameter space for $\theta=2.13^{\circ}$. (a) 
    Direct gap $\widetilde{E}_g$ at the momentum index $24$ sector in the $N=26$ cluster. (b) Direct gap $\widetilde{E}_g$ at the momentum index $15$ sector in the $N=30$ cluster. (c) Gap $E_g$ in the $N=28$ cluster. (d) Schematic phase diagram. The dashed line, which connects the minima of $\widetilde{E}_g$ in the $N=30$ cluster at fixed $\lambda_F$, provides an approximation to the phase boundary between the MR and CDW states.}
    \label{fig:nonabelian-diagram}
\end{figure}

\section{ED Calculation at $\nu_h$=1/2}
\label{appendix:D}
Numerical results at $\nu_h = 1/2$ in tMoTe$_2$ are presented in Fig.~\ref{fig:ED-appendix}(a-c) for a 26 momentum cluster. The energy spectra at $\theta = 2.13^\circ$, $3.15^\circ$, and $3.89^\circ$ exhibit a fourfold quasi-degenerate ground-state manifold. These four quasi-degenerate states occur at momentum indices $6$, $7$, $17$, and $18$, which match with the momentum structure of a composite Fermi liquid in the 0LL at $\nu=1/2$, as shown in Fig.~\ref{fig:ED-appendix}(d). Across all twist angles studied from $2.13^\circ$ to $3.89^\circ$, the 26-momentum cluster spectra provide robust evidence for a composite Fermi liquid at $\nu_h = 1/2$ in tMoTe$_2$. Furthermore, ED on clusters with even numbers of momentum points $N$ from $16$ to $30$ at $\theta = 3.15^\circ$ consistently supports the presence of a composite Fermi liquid, demonstrating the stability of this phase across system sizes. 

\section{ED Calculation at $\theta=2.13^\circ$}
\label{appendix:E}
We present the numerical results at $\theta=2.13^\circ$ and $\nu_h=5/2$. The ED spectra and PES for clusters $N=26,28,$ and $30$ are shown in Fig.~\ref{fig:nonabelian-ED-2.13}. Here we take $\lambda_w=1$ and the wave function  $\wt f_{+,2,\bl{k}}(\bl{r})=[\wt{\psi}_{+,2,\bl k}(\bl r)]^*$ in the projected Hamiltonian $\hat{\mathcal{H}}_2$. For the $N=26$ cluster, a twofold quasi-degenerate ground states  are located  at momentum indices $0$ and $24$, which appears to be consistent with the MR momentum pattern. However, the PES reveals a different nature: instead of the entanglement gap associated with the MR state, it exhibits a clear gap at $\xi_{\mathrm{CDW}}$ and a low-lying level counting of $2\binom{13}{N_A}$, characteristic of a CDW state. A similar situation occurs in the $N=30$ cluster, where a twofold quasi-degeneracy is found at momentum indices $0$ and $15$, while the PES again shows a CDW-like entanglement gap with counting $2\binom{15}{N_A}$. For the $N=28$ cluster, the sixfold MR degeneracy expected at momentum indices $7$, $14$, and $21$ is lifted, and the low-energy states are redistributed across four momentum sectors at $0$, $7$, $14$, and $21$, corresponding to the mBZ center and the midpoints of the mBZ edges. This new momentum pattern is consistent with the CDW state. However, the PES does not show an entanglement gap at $\xi_{\mathrm{CDW}}$, with a low-lying counting of $4\binom{14}{N_A}$ expected for the CDW state, suggesting that the CDW phase is not yet fully developed.

We construct a phase diagram in the $(\lambda_F, \lambda_w)$ parameter space for the three clusters, as shown in Fig.~\ref{fig:nonabelian-diagram}. Here, $\lambda_F \in [0,1]$ controls the weight of the generalized 1LL in the wave function, as defined in Eq.~\eqref{lambdaF}, and $\lambda_w \in [0,1]$ governs the bandwidth, as introduced in Eq.~\eqref{model_nonabelian}. At $\lambda_F = \lambda_w = 0$, the Hamiltonian corresponds to $\hat{\mathcal{H}}_{\mathrm{1gLL}}$ defined in Sec.~\ref{sec:nonAbelian_FCI}, where the wave function is the generalized 1LL obtained via variational mapping, and the bandwidth is set to zero. In this limit, MR states are observed, as characterized by both the ED spectra and the PES for all three clusters. The physical limit occurs at $\lambda_F = \lambda_w = 1$, where the ED spectra and PES  presented in Fig.~\ref{fig:nonabelian-ED-2.13} indicate CDW state.

For the system sizes $N=26$ and $N=30$, the direct gap at momentum index $0$ remains finite throughout the $(\lambda_F, \lambda_w)$ parameter space. We therefore use the direct gap $\widetilde{E}_g$ in momentum sectors $24$ ($15$) for the $N=26$ ($N=30$) cluster to track the phase evolution. For $N=28$ cluster, we use $E_g = E_{7} - E_6$ where $E_i$ is the $i$-th lowest energy level. Along the $\lambda_F = 0$ line, $\widetilde{E}_g$ decreases with increasing $\lambda_w$, reaches a minimum, and then rises again for both $N=26$ and $N=30$ clusters. Similarly, $E_g$ decreases and finally vanishes as $\lambda_w$ increases in the $N=28$ cluster. These behaviors reflect transition between the MR and CDW states: increasing bandwidth destabilizes the MR phase and eventually favors the CDW state. As $\lambda_F$ increases, this trend continues, with the transition point shifting toward larger values of $\lambda_w$. Notably, when the wave function slightly deviates from the generalized 1LL limit ($\lambda_F > 0$), the MR state becomes more robust against finite bandwidth. In the physical limit at $(\lambda_F, \lambda_w) = (1,1)$, the system resides on the CDW side of the transition but can be driven into the MR phase by slightly reducing the bandwidth.

\bibliography{ref}

\end{document}